\documentclass[]{cas-sc}
\usepackage[backend=bibtex,sorting=none,maxcitenames=1,maxbibnames=99,style=numeric-comp]{biblatex}
\usepackage{amsmath,amsthm, amssymb, float, bm, subcaption}
\usepackage{tabularx,booktabs}
\usepackage{comment}
\usepackage{miller}

\usepackage{soul}

\definecolor{C0}{HTML}{1F77B4}
\definecolor{C1}{HTML}{FF7F0E}
\definecolor{C2}{HTML}{2ca02c}
\definecolor{C3}{HTML}{d62728}
\definecolor{C4}{HTML}{9467bd}
\definecolor{C5}{HTML}{8c564b}

\makeatletter
\newlength{\bibsep}{\@listi \global\bibsep\itemsep \global\advance\bibsep by\parsep} % Compatibility with cas-sc
\makeatother
\usepackage[noabbrev,capitalize]{cleveref}
\begin{document}
\shorttitle{}
\shortauthors{}
\title[mode=title]{Facet and energy predictions in grain boundaries: lattice matching and molecular dynamics}
\author[uccs]{Bruno Dobrovolski}
\author[byu-physics]{C. Braxton Owens}
\author[byu-physics]{Gus L. W. Hart}
\author[byu-me]{Eric R. Homer}[orcid=0000-0002-8617-7573]
\author[isu,uccs]{Brandon Runnels}[orcid=0000-0003-3043-5227]
\cormark[1]
\cortext[1]{Corresponding author}
\ead{brunnels@iastate.edu}
\address[uccs]{Department of Mechanical and Aerospace Engineering, University of Colorado, Colorado Springs, CO USA}
\address[byu-physics]{Department of Physics and Astronomy, Brigham Young University, Provo, Utah USA}
\address[byu-me]{Department of Mechanical Engineering, Brigham Young University, Provo, Utah USA}
\address[isu]{Department of Aerospace Engineering, Iowa State University, Ames, IA USA}

\begin{keywords}
  Grain boundaries \\
  Faceting \\
  Computational materials science \\
  Lattice matching \\
\end{keywords}

\begin{abstract}
    Many material properties can be traced back to properties of their grain boundaries.
    Grain boundary energy (GBE), as a result, is a key quantity of interest in the analysis and modeling of microstructure.
    A standard method for calculating grain boundary energy is molecular dynamics (MD); however, on-the-fly MD calculations are not tenable due to the extensive computational time required.
    Lattice matching (LM) is a reduced-order method for estimating GBE quickly; however, it has only been tested against a relatively limited set of data, and does not have a suitable means for assessing error.
    In this work, we use the recently published dataset of \textcite{homer2022examination} to assess the performance of LM over the full range of GB space, and to equip LM with a metric for error estimation.
    LM is used to generate energy estimates, along with predictions of facet morphology, for each of the 7,304 boundaries in the Homer dataset.
    In keeping with prior work, it is observed that LM predictions of low energy boundaries %(``cusps'') 
    matches well with MD results.
    Moreover, there is a good general agreement between LM and MD, and it is apparent that the error scales approximately linearly with the predicted energy value; this makes it possible to establish an empirical estimate on error for future LM calculations.
    An essential part of the LM method is the faceting relaxation, which corrects the expected energy by convexification across the compact space $(S2)$ of boundary plane orientations.
    The original Homer dataset did not allow for faceting, but upon extended annealing, it was shown that facet patterns similar to those predicted by LM were emerging.
\end{abstract}

\maketitle

\section{Introduction}

The interface energy in crystalline materials is influential in a range of applications, such as thermal materials \cite{beyerlein2014emergence}, semiconductors \cite{greuter1990electrical}, large-scale manufacturing 
\cite{raabe2014grain}. 
Specifically, interfaces between like species and phases, i.e., grain boundaries (GB) influence phenomena including short-term failure based on charge transport properties \cite{vladimirov2018energy} and failure in blister steel \cite{raabe2014grain}. 
The study of grain boundaries has existed for many decades, with a resurgence in popularity in recent years due to the increasing importance of GBs at small scales.
Study of grain boundaries dates back to the seminal work of Read and Shockley \cite{read1950dislocation}, with a variety of analytical and numerical models proposed since then.
For a comprehensive overview of grain boundary modeling efforts, the reader is referred to \cite{runnels2016model}.

Grain boundary energy is uniquely tied to the crystallographic structure of the material, with substantially varying results depending on the lattice type \cite{wolf1989correlation, wolf1990correlation, tschopp2007asymmetric, tschopp2007structures, zhao2021structures, merkle1992low, wolf1990structure, wolf1989structure, kang2012atomic, wang2012atomic, wolf1991structure}; although, for simple crystal structures, the GB energy has been observed to be consistent between species up to a scaling factor \cite{bulatov2014grain}.
GBs are often thought to be the {\it least understood} of all material defects due to the enormous space over which their character is defined: three degrees of freedom controlling the orientation relationship between grains, and two more to define the orientation of the interface between them.
Even accounting for crystallographic symmetry, this five dimensional space is difficult to thoroughly explore and parameterize.
Historically, the coincident site lattice (CSL) $\Sigma$ value, defined as the ratio of the CSL unit cell to that of the bicrystal, has been used as a proxy for grain boundary energy.
While the term ``low-$\Sigma$ boundary'' is often used synonymously with ``low-energy boundary'', there is evidence against the notion that low $\Sigma$ is either a sufficient or necessary condition for low energy \cite{Randle:2006:SpecialGBs,EricksonHomer:2020hd}. More importantly,  boundary plane orientation plays a decisive role in the energy of low- and high-$\Sigma$ boundaries alike \cite{Rohrer:2011kp,homer2022examination}.
In fact, it is impossible to properly ascertain the correct boundary energy without full knowledge of all five GB degrees of freedom.

Because of the significant role played by GB energy, it is essential to be able to quickly and accurately determine GB energy for boundaries with arbitrary character.
Experimental methods have been used to obtain estimates of GB energy directly, or to infer energy from GB populations, in a variety of metals and ceramics  \cite{Li:2009rg,Saylor:2004bp,Holm:2011hj,Mishin:2005:NiAl_GBenergy,Beladi:2013,Zhong:2017vd,Randle:2008ig,Dillon:2009ec,Saylor:2003gm}. 
These match well with atomistic simulations \cite{Mishin:2005:NiAl_GBenergy,Barmak:2006:GBenergyAl_exp_sim_comparison,Rohrer:2010df,Holm:2011hj,Barmak:2022:RelativeGBenergiesGBCD}, which provide a reliable means to estimate energy for specific GBs. 
However, atomistic simulations remain a costly method of calculating the energy of a GB with arbitrary character.
Reduced-order models can accelerate our understanding and the development of materials.
A variety of models have been proposed, ranging from the original model of Read and Shockley for low angle boundaries \cite{read1950dislocation}, to machine learning models based on MD datasets \cite{pinz2022data,rosenbrock2017discovering}.

The lattice matching (LM) approach for GB energy is a semi-analytical method that is able to rapidly calculate grain boundary energy for arbitrary crystals and orientations.
LM has been shown to accurately reproduce energy cusps for a wide range of boundaries \cite{runnels2016analytical}, and has been successfully used in mesoscale simulations \cite{gokuli2021multiphase}.
One of the key advantages of LM is the minimal number of parameters required: other than the crystal structure and bicrystal configuration, LM relies on only three model parameters, one of which is a scaling constant.
The ability of LM to calculate energy accurately with a minimal number of model parameters builds confidence in its ability to predict GB energy in regions where little or no information is available from MD or experiment.
LM has also been used in conjunction with grain boundary relaxation to simultaneously improve the LM energy estimate, while providing predictions about faceting behavior in the boundary.
It has been shown that, even if faceting does not actually occur, the relaxation method substantially improves the LM method \cite{runnels2016relaxation}.
In other cases, LM has been shown to correctly predict microfacets in agreement with experimental observation  \cite{leff2019determination}.

A perennial limitation in GB modeling has been the lack of available atomistic data for comparison.
The survey of \textcite{olmsted2009survey} has been used for nearly 15 years, but is not exhaustive enough to ensure sufficient and uniform coverage over GB space.
Recent work by Homer {\it et al} has resulted in a comprehensive GB dataset that semi-uniformly spans GB space using a single potential and consistent methodology \cite{homer2022examination,Homer:2022:AlGBdataset}.
7,304 unique aluminum GBs were selected to provide comprehensive coverage the full 5D disorientation space. 
More than 43 million GB structures were considered to find the minimum energy for each GB as well as the corresponding atomic structure.
Because this work focused on retaining near-planar boundaries; the GB construction process only used conjugate gradient minimization and did not heat the atomic structure, thereby minimizing the amount of faceting that occurred.
(Note: this does not mean the structures didn't facet at all, just that it was not promoted.)
This kind of dataset is therefore better suited for testing the LM method and providing a reliable error estimate for LM predictions.

The contribution of this work is twofold.
First, the overall performance of LM (with and without faceting relaxation) is examined over the full range of GB space, and compared to the energies reported in \cite{homer2022examination,Homer:2022:AlGBdataset}.
This comparison allows the reliability of LM to be established, along with estimates of error in LM predictions.
Second, facet patterns as predicted by LM are considered, in order to determine the likelihood of faceting and the expected impact on the energy estimation.
The LM facet patterns are compared to the boundary morphologies in the MD dataset, as well as some new annealed MD simulations.

The remainder of the paper is structured in the following way.
\Cref{sec:methods} provides an overview of the MD and LM methodologies and algorithms.
The LM data is presented in \cref{sec:results}, and compared to the MD results (from the dataset and newly computed annealed structures) for grain boundary energy, error estimation, and facet morphology.
\Cref{sec:discussion} contains a discussion on some of the insights and perspectives of the two GB modeling approaches.

%%% Local Variables: 
%%% mode: latex
%%% TeX-master: "main.tex"
%%% End: 

\section{Methods}\label{sec:methods}

This work aims to synthesize the lattice-matching model with a very large MD dataset of grain boundary energies and structures.
Here, we provide a brief overview of the methods used in this work: MD, LM, facet relaxation, and thermalization of selected MD boundaries to encourage faceting. 

\subsection{Generation of molecular dynamics dataset}\label{sec:md_generation}

The creation of the computed aluminum grain boundary dataset used in this work is described in detail in \cite{homer2022examination} and is available for download \cite{Homer:2022:AlGBdataset}. 
In short, coincidence site lattice (CSL) theory was used to select the points in the 5D disorientation space to provide comprehensive, but not uniform, coverage. 
With the crystallographic character of the 7304 grain boundaries selected, an average of 5888 potential starting structures were created for each GB to find the minimum energy structure for each boundary. 
The starting structures were obtained by varying the relative placement of the two crystals, the placement of the boundary plane, and allowed proximity of atoms in the grain boundary region. 
Each of these starting structures were then subjected to conjugate gradient minimization, without heating the structure first, and the lowest energy structure was kept as the representative GB structure. 
Trends in the dataset are examined in \cite{homer2022examination}. 
In the present work, these GB energy values are used for comparison with predictions by LM.

\subsection{Lattice matching method for grain boundary energy}
Here we review the LM method for grain boundary energy.
A general reference for this section is \cite{runnels2016analytical}.
LM is rooted in the construction of optimal transportation theory, building on the assumption that the grain boundary energy is proportional to the energetic cost of transforming one lattice into another within the interfacial region.
The grain boundary energy, then, is said to be the minimizer of this cost, making the problem one of optimal transport.
Applying thermalization and taking the $L^2$ upper bound makes it possible to determine the solution in terms of lattice density fields.
The lattice density field for a grain, $\bm{\rho}$, is defined implicitly as
\begin{align}
    \int_\Omega \rho(\bm{x})\,d\bm{x} = \langle N \rangle_\Omega, \ \ \ \forall\ \mathrm{ measurable }\ \Omega \subset \mathbb{R}^3,
\end{align}
where $\langle N \rangle_\Omega$ is the expected value of the number of atoms contained within $\Omega$.
Intuitively one may think of $\rho(\bm{x})$ as a collection of distributions (often, Gaussians), each centered at the location of an atom in the lattice.
Lattice density functions are represented using Fourier series,
\begin{align}
    \rho(\bm{x}) = \sum_{\bm{n}\in\mathbb{Z}^3} e^{i\bm{k}_{\bm{n}}\cdot\bm{x}} \, \hat\psi(\bm{k}_{\bm{n}}) \, \rho_{\bm{n}},
\end{align}
where $\bm{k}(\bm{n})$ are the reciprocal lattice vectors, $\hat{\psi}$ is the reciprocal atomic distribution (often, a Gaussian), and $\rho_{\bm{n}}$ are Fourier coefficients.
The energy is then proportional to the normalized covariance of the two atomic distributions over the boundary plane:
\begin{align}
    c[\rho^1,\rho^2] = \frac{1}{\mu(\bm{1})}\int_{\partial\Omega} \rho^1(\bm{x})\,\rho^2(\bm{x})\,d\mu(\bm{x})    
\end{align}
where $\mu$ is a measure with respect to a window function $\phi$, the selection of which is determined heuristically.
If specializing to two identical lattices undergoing rotations $\bm{R}^1,\bm{R}^2\in SO(3)$, then the covariance is expressible in closed form,
\begin{align}\label{eq:covariance}
    c[\rho^1,\rho^2](\bm{R}^1,\bm{R}^2) = \frac{1}{\hat\phi(\bm{0})} \sum_{\bm{n}^1,\bm{n}^2\in\mathbb{Z}^3}
    \rho^1_{\bm{n}^1}\,\rho^2_{\bm{n}^2}
    \hat\psi(\bm{k}_{\bm{n}^1})
    \hat\psi(\bm{k}_{\bm{n}^2})
    \hat\phi(\mathcal{P}(\bm{R}^1\bm{k}_{\bm{n}^1} - \bm{R}^2\bm{k}_{\bm{n}^2})),
\end{align}
where $\hat\phi$ is the Fourier transform of $\phi$ and $\mathcal{P}$ is a projection operator onto the boundary plane.
\Cref{eq:covariance} is a double sum, closed-form expression that is readily amenable to evaluation.
The Fourier coefficients $\rho^1,\rho^2$ can be precalculated in advance, increasing the efficiency of the calculation.
The convergence of the double sum is dependent upon the thermalization parameter, and tends towards infinity as the ``temperature'' of the lattices tends towards zero.
For most practical applications, a relatively high value of the thermalization parameter is sufficient; nevertheless, the double sum often requires thousands of terms to guarantee convergence.
Therefore, some care must be taken when implementing \cref{eq:covariance} to avoid excessive computation time.
We find that a C++ implementation with O3 optimization is able to yield a GB energy calculation on the order of milliseconds.
The reader is referred to WIELD \cite{wield}, which contains an optimized C++ implementation and Python bindings, for further details.

To convert from covariance to grain boundary energy $\gamma$, the following affine relationship is used, which we consider to be a function of rotations only (i.e., treating the lattice fields as parameters):
\begin{align}\label{eq:lm}
    \gamma(\bm{\mathrm{R}}^1,\bm{\mathrm{R}}^2) = \gamma_0\Big(1 - \frac{c(\bm{\mathrm{R}}^1,\bm{\mathrm{R}}^2)}{c_0}\Big).
\end{align}
The value $c_0$ is the ``ground state'' covariance, and corresponds to the maximum possible value of $c_0$ attainable for the given lattices.
This value can be calculated directly for any pair of lattices by evaluating the no-boundary boundary along the lattice's most densely packed plane, and so this is not considered to be an adjustable parameter.

The value $\gamma_0$, however, is a scaling factor that must be determined heuristically by comparison to experimental, atomistic, or {\it ab initio} calculations.
Values for $\gamma_0$ were previously determined by comparison to a consistent set of atomistic data \cite{runnels2016analytical}.
However, atomistic results are known to vary in their calculation of grain boundary energies by a multiplicative factor \cite{wolf1991structure}, which may produce error in the calculation of $\gamma_0$.
Moreover, grain boundary calculations exhibit strong potential dependence,
Calculations can differ by a multiplicative factor up to an order of magnitude depending on the model used, although they do demonstrate remarkable consistency (e.g., in cusp location and shape) with themselves and LM, up to scalar multiplication \cite{waters2023automated}.
For aluminum specifically, the selection of potential can yield a difference in scaling factor up to four; see \cite{TE918853243284003,TE202986963854001,TE117904176283001,TE641102822364001} for the complete data set including comparisons for the potential considered here.

Given the variability in atomistic data, it becomes necessary to quantify the uncertainty in $\gamma_0$ in order to ensure reliability of future grain boundary energy predictions.
Determination of uncertainty is one of the objectives of this work.

\subsection{Faceting relaxation with lattice matching}

As indicated above, many grain boundaries exhibit a faceted structure\footnote{Specifically, this is called ``micro-faceting,'' which is distinct from the formation of large, sometimes even visible, piecewise planar sections of boundary. In this work, ``faceting'' refers exclusively to micro-faceting.}.
Mathematically, this is understood to be a consequence of the extreme non-convexity of $\gamma$ with respect to boundary orientation \cite{bulatov2014grain}.
Faceting is relatively easy to see and measure experimentally, providing a basis for implicit validation of energy.
Facet morphology is, however, impervious to the choice of scaling constant $\gamma_0$ \cite{runnels2016relaxation}, providing little help for experimental validation of $\gamma_0$.

In lattice-matching calculations, it has been shown that straightforward calculation of \cref{eq:lm} will systematically overestimate grain boundary energy.
This can be seen by observing that evaluation of $\gamma$ for no-boundary boundaries (i.e., $\gamma(\bm{\mathrm{R}},\bm{\mathrm{R}})$) will produce a nonzero value when the boundary plane does not align with the densest packed plane.
The faceting algorithm, when applied to LM, is able to resolve this discrepancy by constructing a real (or, sometimes, hypothetical) faceted boundary from nearby energy cusps.
Formally, the process for determining the facet-relaxed energy $\gamma_C$ is 
\begin{align}\label{eq:relaxation}
    \gamma_{NC}(\bm{n}) = \inf_{\substack{\bm{n}^i\in S(2)\\\lambda^i\in\mathbb{R}^+}} \sum_{i=1}^N \lambda^i\,\gamma(\bm{n}^i)
    \ \ \ \text{ subject to } \ \ \ 
    \sum_{i=1}^N \lambda^i \, \bm{n}^i = \bm{n}, \ \ \ 
    \sum_{i=1}^N\lambda^i\ge 1,
\end{align}
where $\bm{n}^i$ are normal vectors corresponding to facet plane orientations, $\lambda^i$ are area fractions corresponding to each facet, and $N$ is the facet order.
$N=1$ corresponds to the no-facet case.
$N=2$, referred to as ``two-faceting'' or ``second-order'' faceting, is the commonly observed zig-zag facet pattern, extruded in the direction mutually orthogonal to both facets.
$N=3$ is referred to as ``three-faceting'' or ``third-order'' faceting, and corresponds to a three-dimensional facet pattern.
Higher order facet patterns ($N\ge 4$) may exist, but cannot lower the energy beyond what is possible with third order faceting relaxation (see \cite{runnels2016relaxation} for discussion), and come at the expense of additional facet junction energy. 
This does not mean that they do not exist, only that they do not offer an energetic advantage from a LM persepective. 
Note that the inequality in \cref{eq:relaxation} reflectsthat the process of faceting corresponds to a net increase in area.

There is, to our knowledge, no way to evaluate \cref{eq:relaxation} in closed form, due to the nonconvexity of $\gamma$.
Fortunately, $S(2)^N$ is a compact space and, depending on the orientation relationship, can often be reduced by a factor of up to 8.
Therefore, we found that a primitive grid search is sufficient for solving \cref{eq:relaxation} without incurring excessive computation time.
We used the WIELD convexification algorithm to calculate $\gamma_{2C}$ and $\gamma_{3C}$ for each boundary, and found that even boundaries with irreducible fundamental zone took only on the order of seconds to minutes to compute.

In this work, we use this algorithm to generate second and third order faceting predictions for LM.
Faceting, when applied to LM, does not always mean that a visible facet pattern will emerge, since it sometimes functions merely as a regularization.
On the other hand, LM has been shown to predict experimentally and atomistically observed facet patterns \cite{leff2019determination,runnels2016relaxation}.
We remark that the dataset presented in \cite{homer2022examination} specifically aimed to suppress faceting in order to explore the energies for planar or near-planar boundaries.
Thus, one aspect of the comparison performed in this work is the assessment of when the predicted faceting of LM functions as a morphological prediction, and when it functions merely as an energy correction.
In section \cref{sec:results_faceting}, we explore the predictive capability of the faceting algorithm by testing to see if facet-likely atomistic boundaries do form boundaries upon sufficient annealing time.

\subsection{Observation of faceting through annealing}

As noted in the introduction, the amount of faceting in the original MD dataset (\cref{sec:md_generation}) was minimized without thermalizing the GBs, or heating the GBs to high temperatures, prior to the minimization process. 
This was to deliberately suppress the formation of facets, because faceted boundaries consist of boundary plane normals with energies that are different than that of the overall boundary plane normal.
So, working with non-thermalized GBs made it possible to determine the most likely energy for a given boundary plane prior to faceting. 
With comprehensive information about the energy as a function of boundary plane, one can then {\it predict} likely faceting that could lower the energy of a given boundary. 

Since the LM relaxation algorithm can predict likely faceting to reduce the overall energy of a GB, the non-thermalized MD dataset provides an ideal opportunity for testing the predictive capability of the LM relaxation algorithm:
A GB that has had a chance to facet will, ostensibly, adopt the faceted structure predicted by the LM model. 
Therefore, the non-thermalized MD boundaries were selected that had the highest likelihood, as predicted by LM, of faceting.
Because of the strong tendency of $\Sigma3$ GBs to facet \cite{Abdeljawad:2018hz}, all $\Sigma3$ GBs are examined to see how their structure would evolve given thermal energy to facilitate the process. 
To accomplish this using MD, each of the $\Sigma3$ GBs are replicated by 3 in both directions of the GB plane to increase the GB area by a factor of 9. 
This makes it easier for long facets to emerge within the constraints of the periodic boundary conditions. 
The GBs are then equilibrated at 750 K for 50 ps, then quenched with a first order (exponential) cooling profile using a time constant of 4.34 ps over 30 ps. %T(t)=750*exp(-t/4.34)
Finally, the structure is minimized using conjugate gradient minimization. These ``annealed'' GB structures are compared with the faceting predictions by the LM model.

%%% Local Variables: 
%%% mode: latex
%%% TeX-master: "main.tex"
%%% End: 

\section{Results}\label{sec:results}

Given the approach outlined in the previous section, the goal of this section is to determine three things: (1) what is the expected error of LM as determined by systematic comparison to atomistics; (2) are the cusps predicted by LM qualitatively similar to those predicted by atomistics; and (3) what is the relationship between LM facet predictions and original/annealed molecular dynamics?
All LM results, including the unrelaxed energy, relaxed energy by 2-faceting, relaxed energy by 3-faceting, and associated facet orientations, are included as supplementary data\footnote{\url{Al_GBs_2022_LatticeMatching.csv}}.
Boundary identifiers are consistent with those used in the MD dataset by \textcite{homer2022examination}, which is available for download \cite{Homer:2022:AlGBdataset} .

\subsection{Grain boundary energy}\label{sec:grain_boundary_energy}

\begin{figure}
  \includegraphics[width=\linewidth]{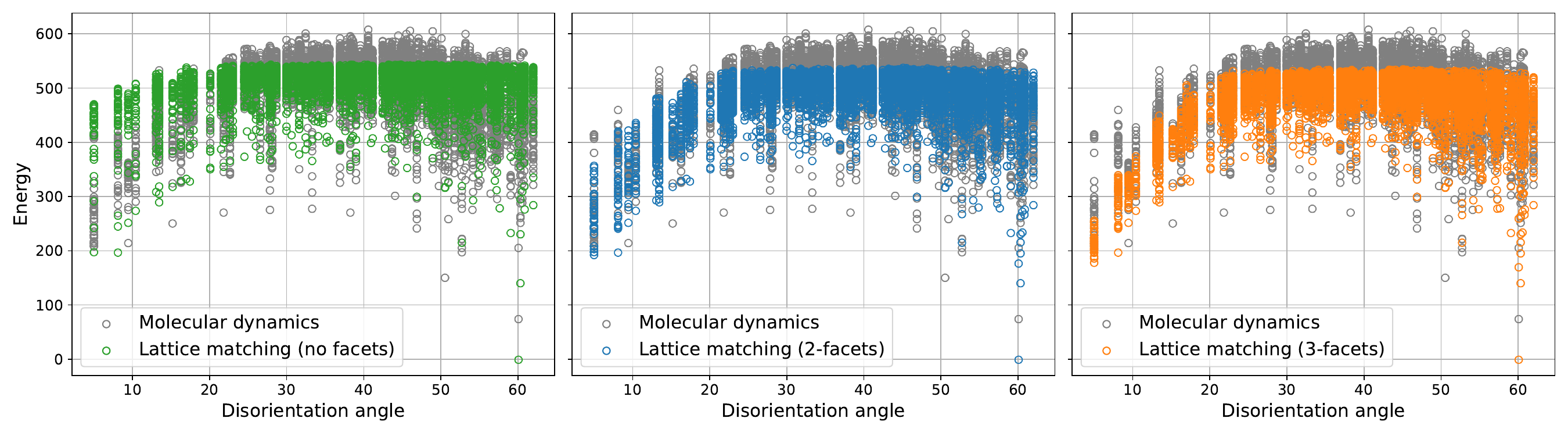}
    \caption{
        Comparison of LM to MD data, plotted with respect to disorientation angle. 
        (Left, green) Unrelaxed LM data
        (Center, blue) LM relaxed with second order (1d) faceting
        (Right, gold) LM relaxed with third order (2d) faceting
        (All, gray) MD data
    }
    \label{fig:disorientation}
\end{figure}
\begin{figure}
  \includegraphics[width=\linewidth]{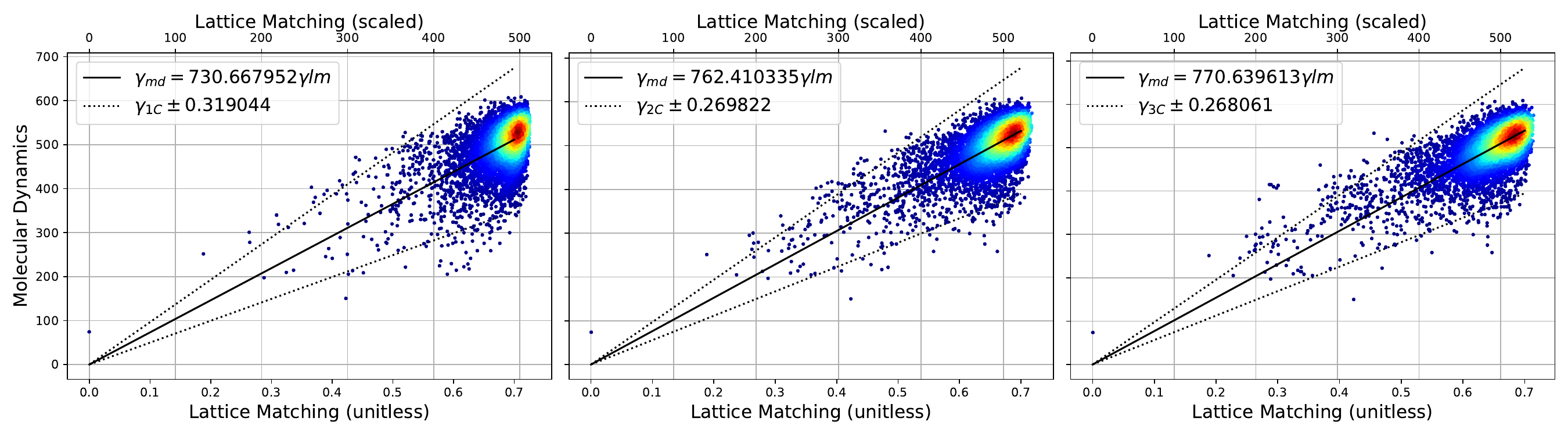}
  \caption{Comparison between LM and MD data for unrelaxed (left), second order facet relaxed (center), and third order facet relaxed (left).
    Abscissa is LM value, ordinate is MD value; the black line indicates a perfect match, and the dotted lines indicate approximate error.
    Color indicates density of points.
  }
  \label{fig:parityplot}
\end{figure}

We begin by presenting the raw LM energy prediction with the energy prediction from atomistic simulations as a function of disorientation angle (\cref{fig:disorientation}).
The left, center, and right panels correspond to the no faceting case, 1-facet case, and 2-facet case, respectively.
The first qualitative trend to notice is the similar behavior for low angle boundaries, where the disorientation angle is less than $20^\circ$.
This is reflective of the well-known ability of LM to accurately capture cusp locations.
For the same reason, similar behavior is again observed for the lower energy grain boundaries as the disorientation goes to $60^\circ$.

One striking feature is the presence of a clear upper bound in the LM data, represented by a plateau above which there are no LM points.
This is well-known behavior, and corresponds to the fact that LM has a finite upper limit corresponding to the no-coincidence case.
On the other hand, a significant portion of the atomistic data exceeds this value.
There are multiple possible reasons for this, with the most straightforward being that LM simply is not valid for certain high angle boundaries.
A contributing factor may be that the atomistic simulations did not actually find the minimum energy state of the boundary, but instead arrived at a metastable, higher energy state. 
However, given the extensive optimization performed in the original MD dataset, we believe this to be unlikely.
And, though we cannot be be sure that the atomistic simulations did achieve the minimum energy configuration, it is not expected that they would exhibit the type of upper bound exhibited by the LM results.

The same data is presented again in a parity plot (\cref{fig:parityplot}), where each point signifies a particular boundary; its horizontal position represents its LM energy and its vertical position represents its atomistic energy.
Perfect agreement between LM and atomistics would result in all points falling along a straight line, the slope of which is the scaling parameter $\gamma_0$.
Obviously, from the figure, there is substantial variance between the two models.
This is not surprising, as LM is not generally expected to capture large angle boundary energy perfectly.
There is a linear trend between the two energy calculations.
Least squares regression is used to determine the most likely scaling factor, which was $731,762,770$, for the no-facet, two-facet, and three-facet cases, respectively.
The standard deviation was also calculated and reported to be $32\%,27\%,27\%$ for each case, and the corresponding dispersion estimate is indicated by dotted lines above and below the mean trendline.
Nearly all of the points fall within this band, with only a handful outside.
Therefore, from this regression analysis, that the maximum error anticipated for LM is $\pm25\%$.
Given that these data were obtained using only three parameters, we find this error to be reasonable.

It is important to note the influence of faceting on the scaling factor and standard deviation.
Apparently, second order faceting results in a substantial decrease in the error, while third order faceting does not narrow the window by any significant amount.
Thus, it is concluded that second order faceting is sufficient when performing actual energy calculations (which is fortunate, because of the reduced time needed for second order vs. third order faceting.)
Still, third order faceting produces an interesting change in the distribution of points, because even though the error is unchanged, there is a noticeable difference in the scatter plots.
One small cluster of points, in particular, located in the three-facet plot at approximately $(0.29,400)$, is not present in the two-facet or no-facet cases.
This implies that these points experienced a substantial energy difference, dropping the energy well below the atomistic estimate.
This is likely due to the constraints placed on the atomistic simulations that discourage faceting.

\subsection{Exploration of grain boundary energy in the fundamental zone} \label{sec:cusps}

%4cm 1.5cm 4cm 1.5cm
\def\tL{2}
\def\tB{1.6}
\def\tR{3.9}
\def\tT{1.5}
\def\height{2.98}
\begin{figure}
  \begin{subfigure}{0.24\linewidth}\includegraphics[height=\height cm,clip,trim=\tL cm \tB cm \tR cm \tT cm]{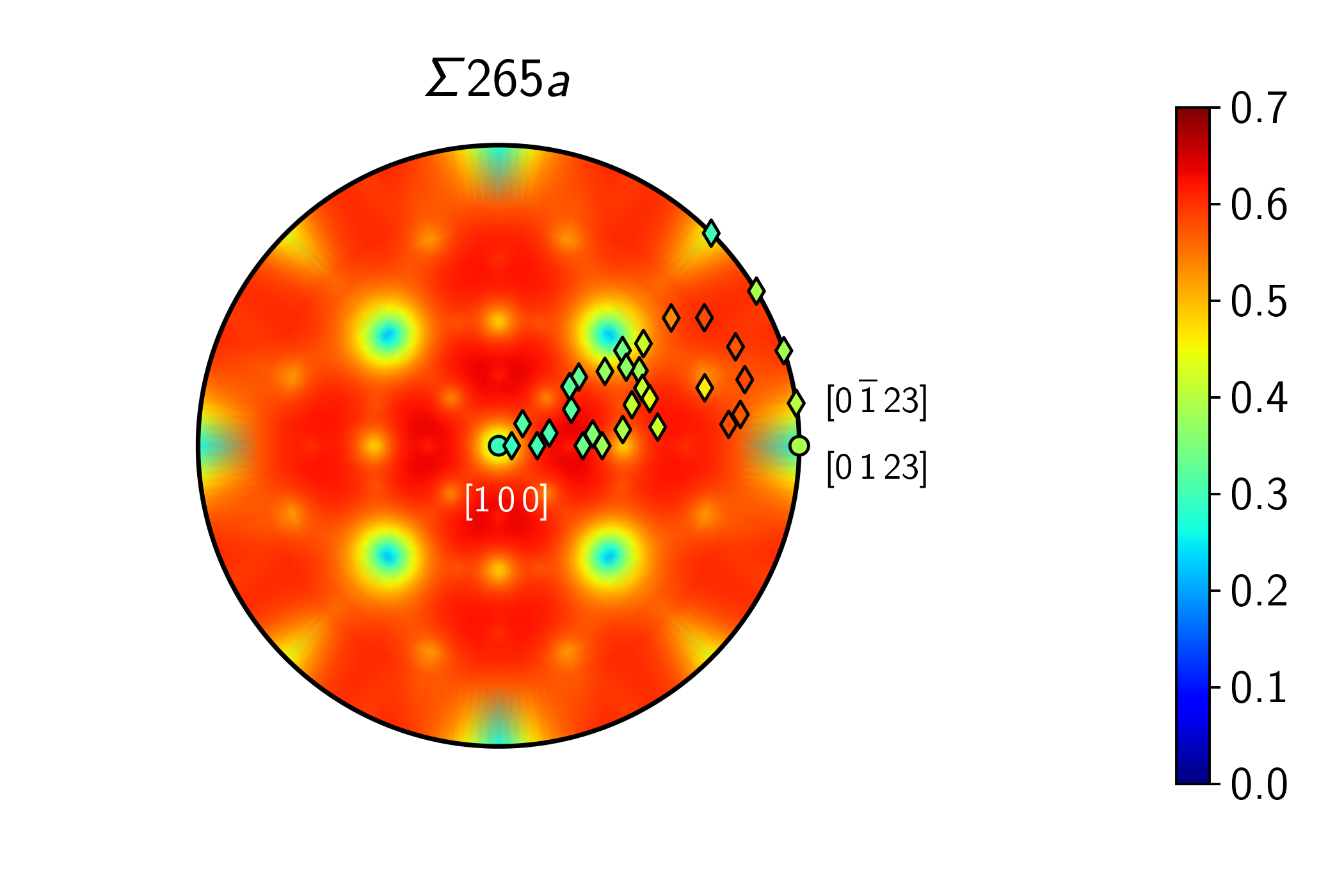}\caption{$\Sigma265a - 5.0^{\circ} [1\,0\,0] - D_{4h}$}\label{fig:facet_s265a}\end{subfigure}%
  \begin{subfigure}{0.24\linewidth}\includegraphics[height=\height cm,clip,trim=\tL cm \tB cm \tR cm \tT cm]{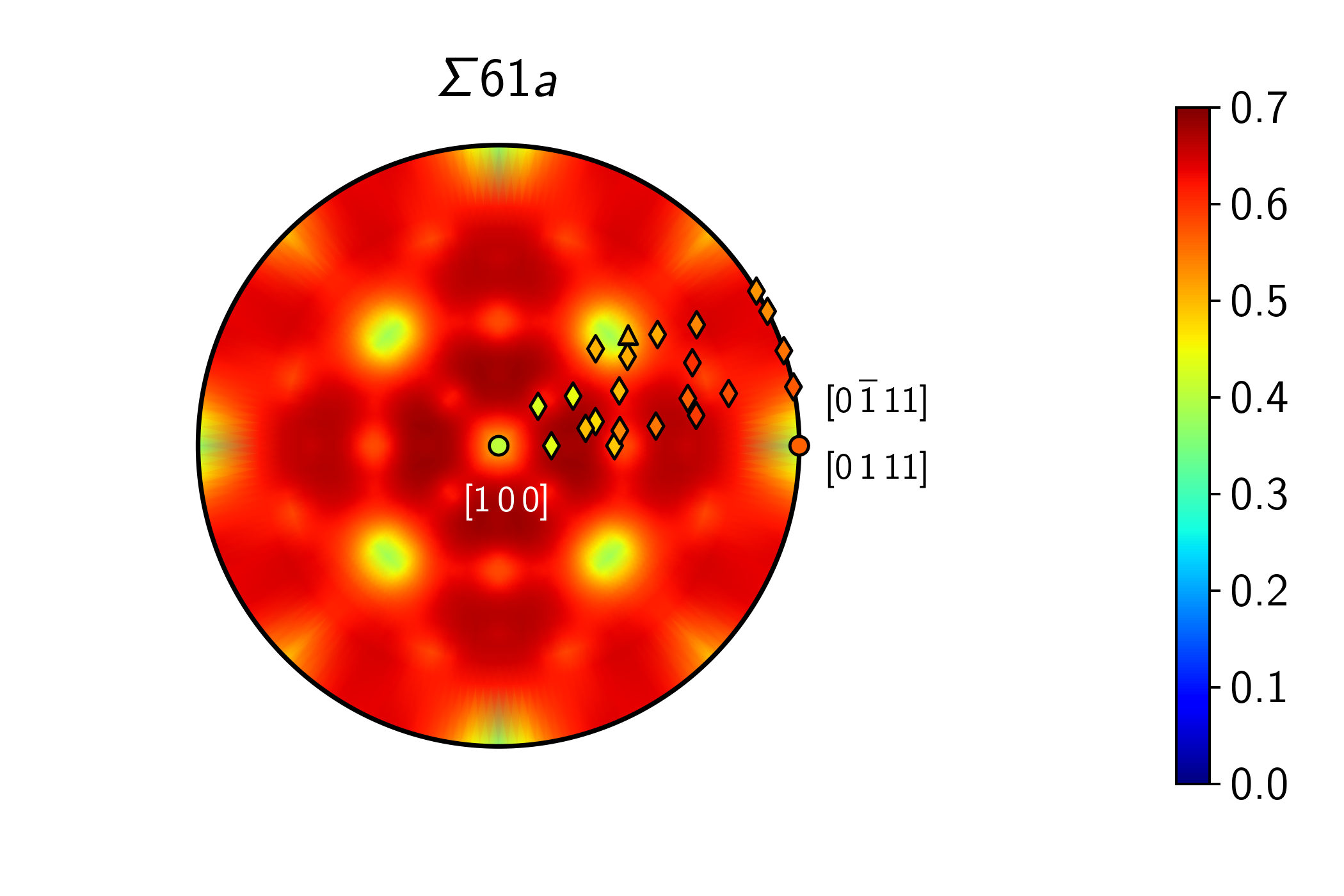} \caption{$\Sigma61a - 10.4^{\circ} [1\,0\,0] - D_{4h}$}\label{fig:facet_s61a}\end{subfigure}%
  \begin{subfigure}{0.24\linewidth}\includegraphics[height=\height cm,clip,trim=\tL cm \tB cm \tR cm \tT cm]{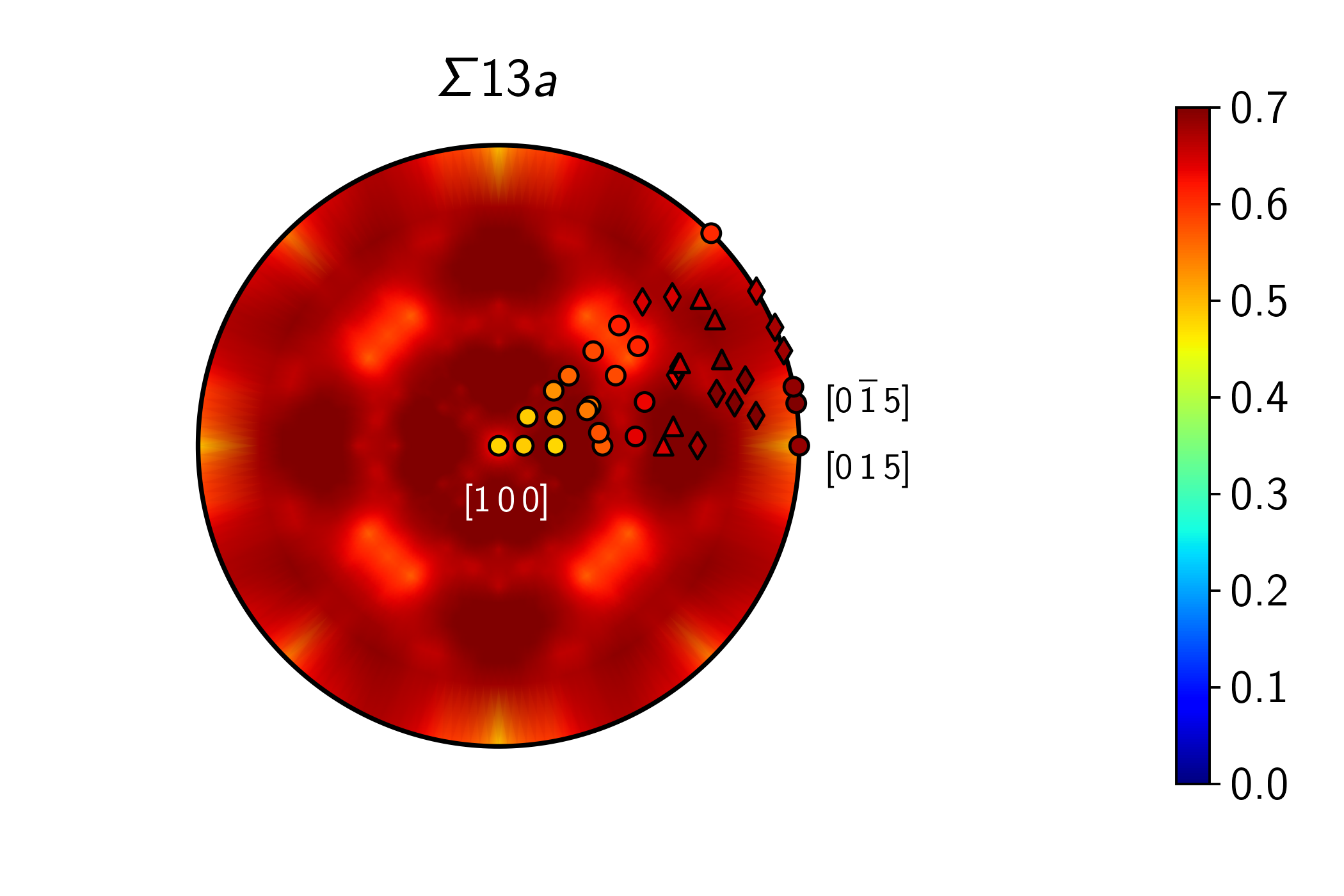} \caption{$\Sigma13a - 22.6^{\circ} [1\,0\,0] - D_{4h}$}\label{fig:facet_s13a}\end{subfigure}%
  \begin{subfigure}{0.24\linewidth}\includegraphics[height=\height cm,clip,trim=\tL cm \tB cm \tR cm \tT cm]{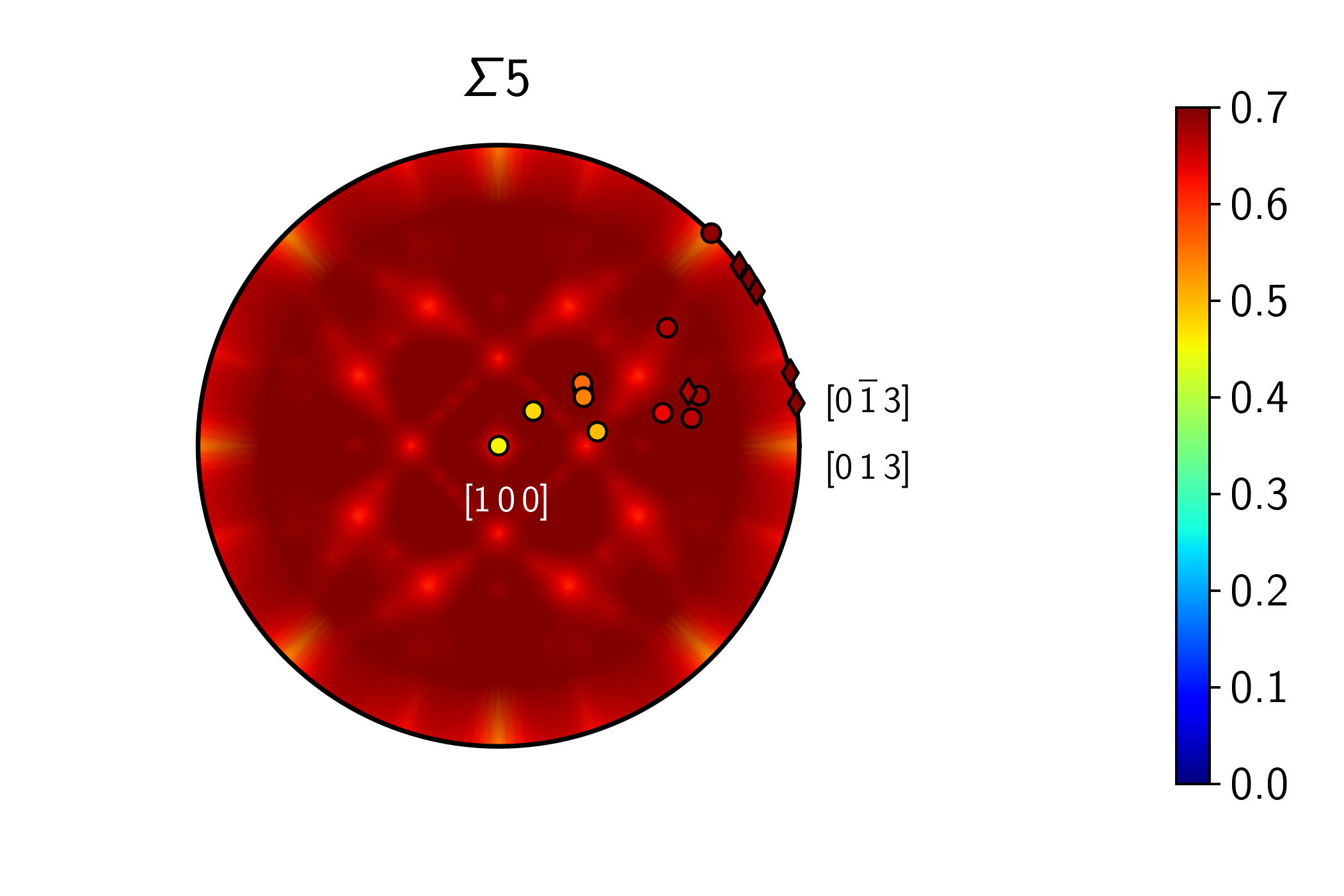} \caption{$\Sigma5 - 36.9^{\circ} [1\,0\,0] - D_{4h}$}\label{fig:facet_s5}\end{subfigure}%
  \begin{subfigure}{0.08\linewidth}\includegraphics[height=\height cm,clip,trim=13cm 0.5cm 0cm 1cm]{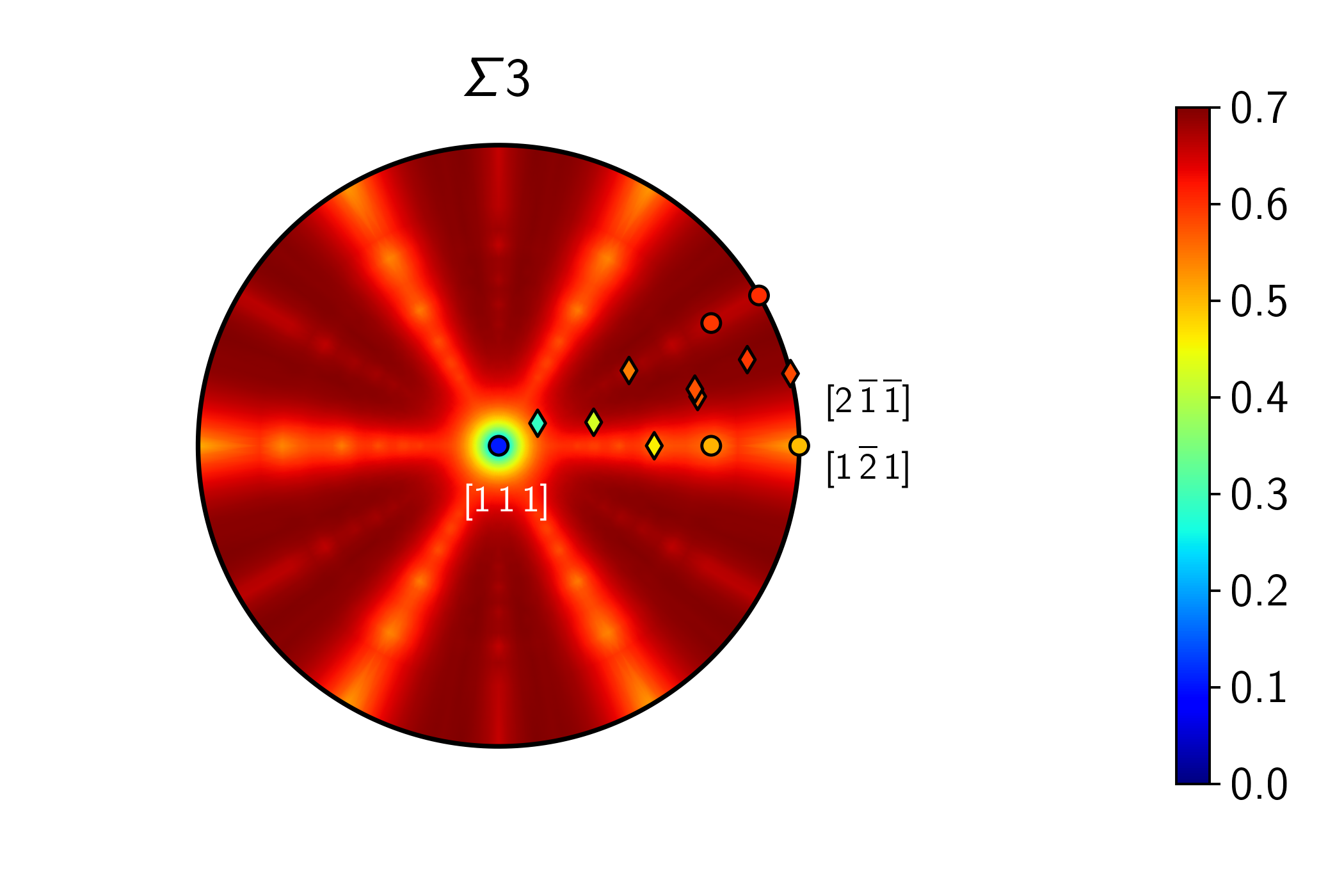}\end{subfigure}

  \begin{subfigure}{0.24\linewidth}\includegraphics[height=\height cm,clip,trim=\tL cm \tB cm \tR cm \tT cm]{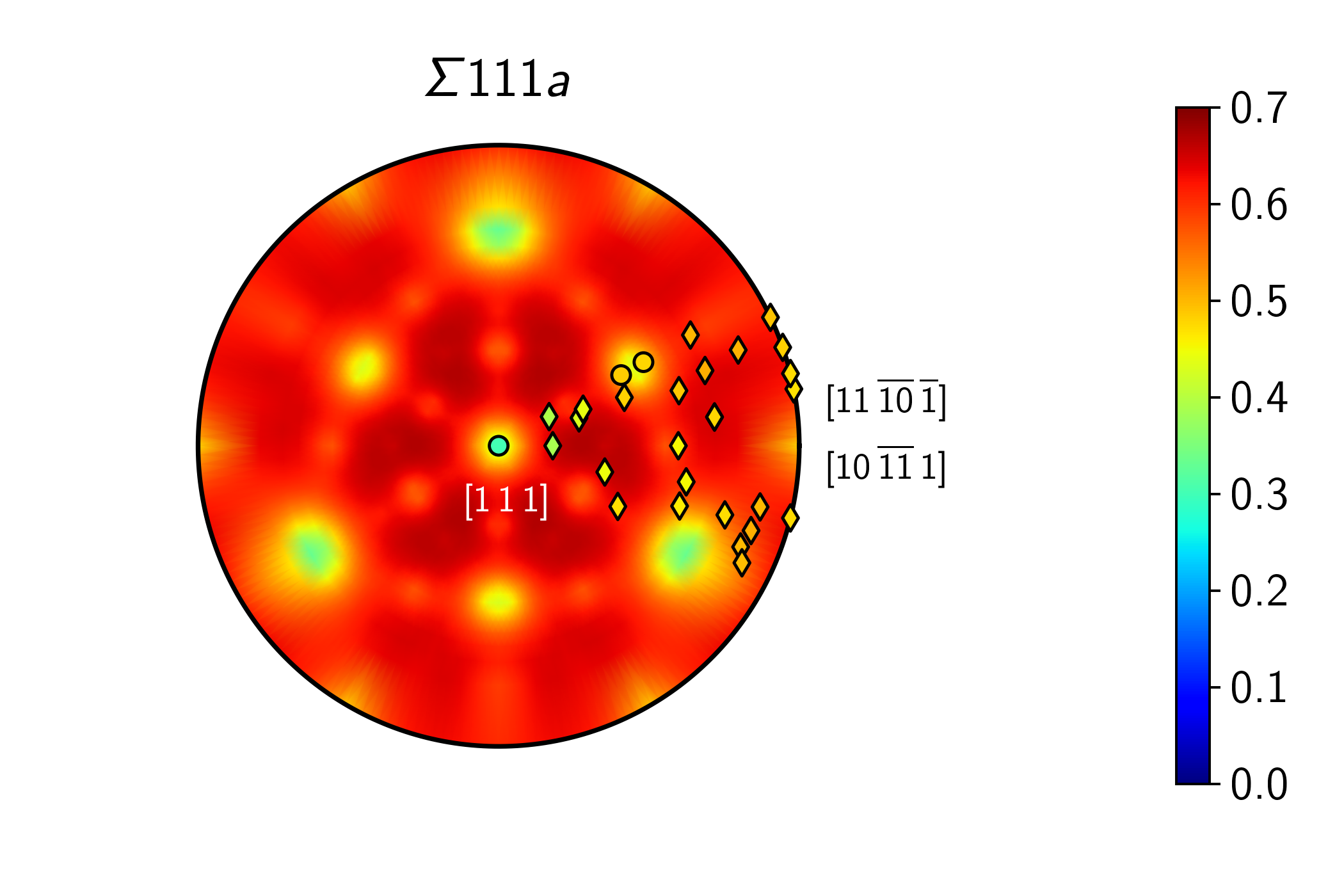} \caption{$\Sigma111a - 9.4^{\circ} [1\,1\,1] - D_{3d}$}\label{fig:facet_s111a}\end{subfigure}%
  \begin{subfigure}{0.24\linewidth}\includegraphics[height=\height cm,clip,trim=\tL cm \tB cm \tR cm \tT cm]{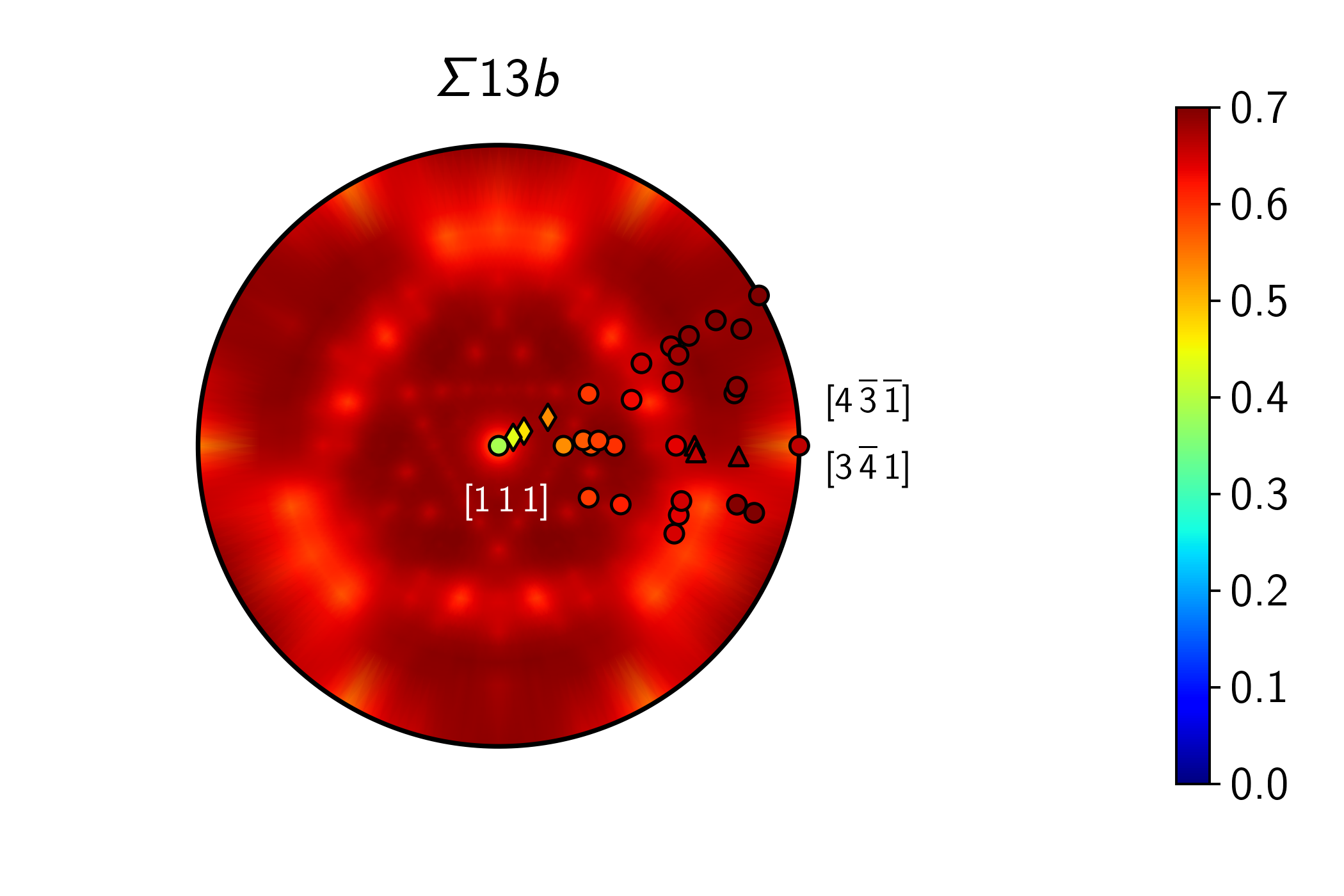} \caption{$\Sigma13b - 27.8^{\circ} [1\,1\,1] - D_{3d}$}\label{fig:facet_s13b}\end{subfigure}%
  \begin{subfigure}{0.24\linewidth}\includegraphics[height=\height cm,clip,trim=\tL cm \tB cm \tR cm \tT cm]{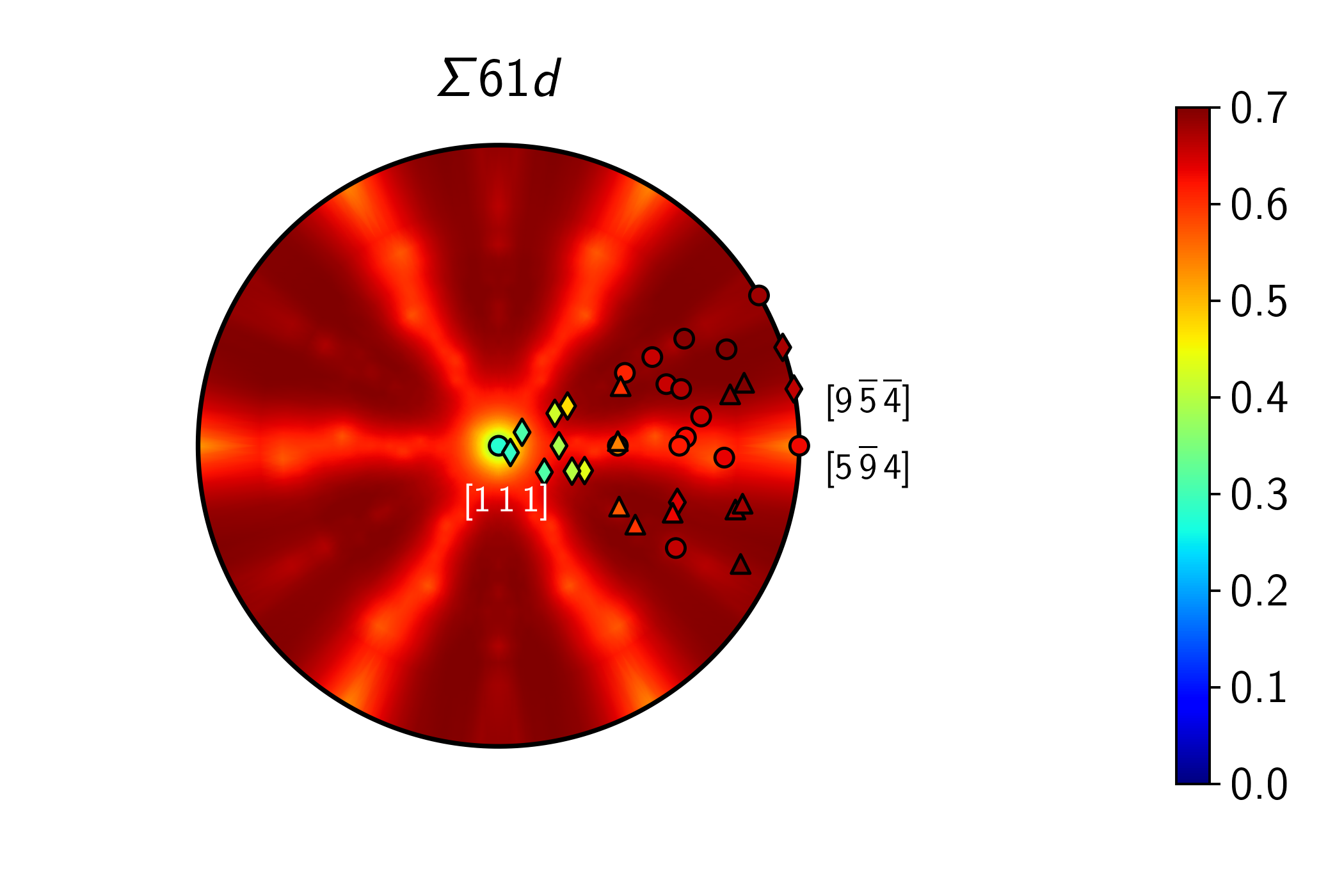} \caption{$\Sigma61d - 52.7^{\circ} [1\,1\,1] - D_{3d}$}\label{fig:facet_s61d}\end{subfigure}%
  \begin{subfigure}{0.24\linewidth}\includegraphics[height=\height cm,clip,trim=\tL cm \tB cm \tR cm \tT cm]{results/bpfz/3.png} \caption{$\Sigma3 - 60.0^{\circ} [1\,1\,1] - D_{6h}$}\label{fig:facet_s3}\end{subfigure}%
  \begin{subfigure}{0.04\linewidth} \ 
  \end{subfigure}

%alternates, not used
% 131e - 60.252 [5 5 4]
% 67d - 60.493 [4 3 3]
  
  \begin{subfigure}{0.24\linewidth}\includegraphics[height=\height cm,clip,trim=\tL cm \tB cm \tR cm \tT cm]{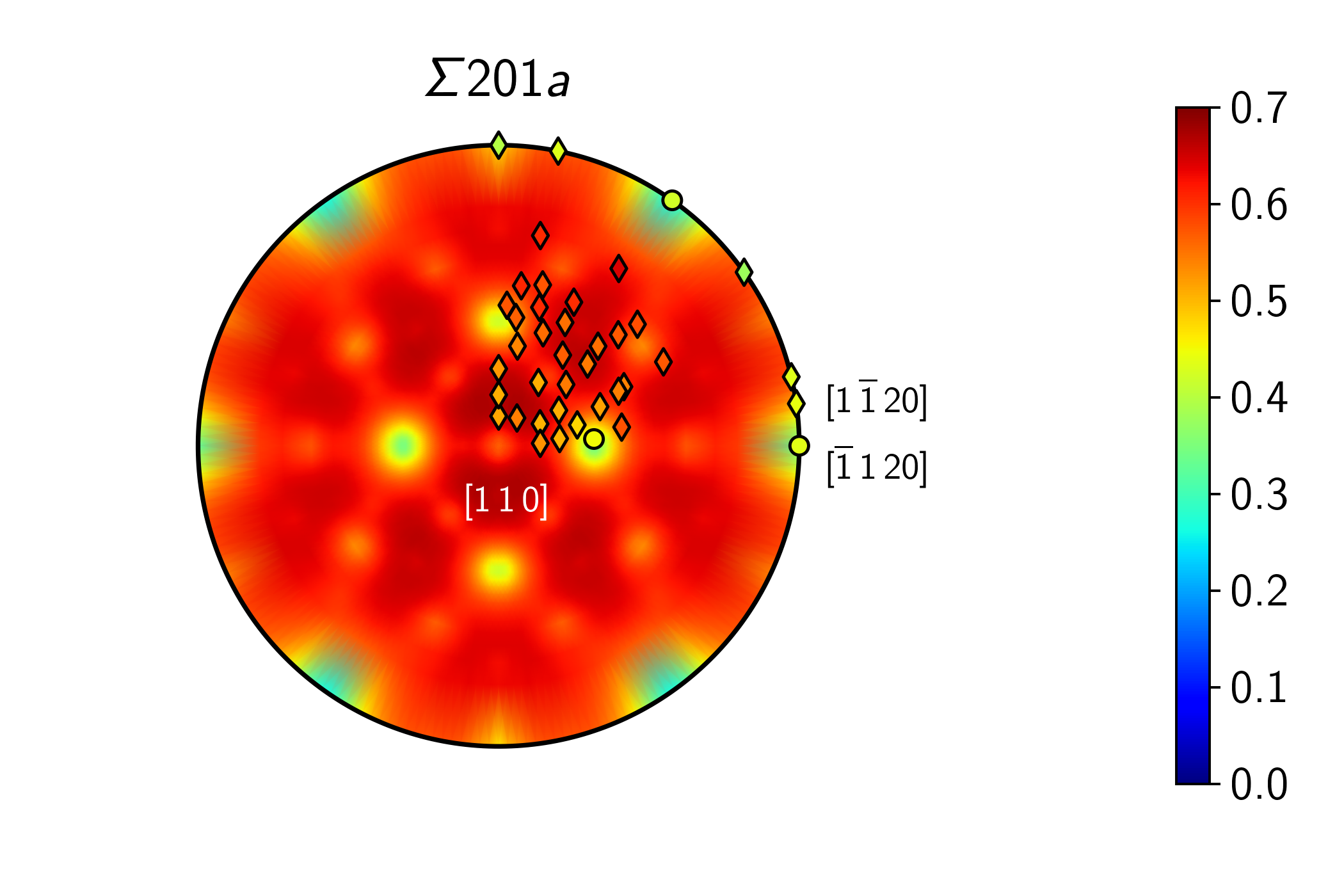}
  \caption{$\Sigma201a - 8.1^{\circ} [1\,1\,0] - D_{2h}$}\label{fig:facet_s201a}\end{subfigure}%
  \begin{subfigure}{0.24\linewidth}\includegraphics[height=\height cm,clip,trim=\tL cm \tB cm \tR cm \tT cm]{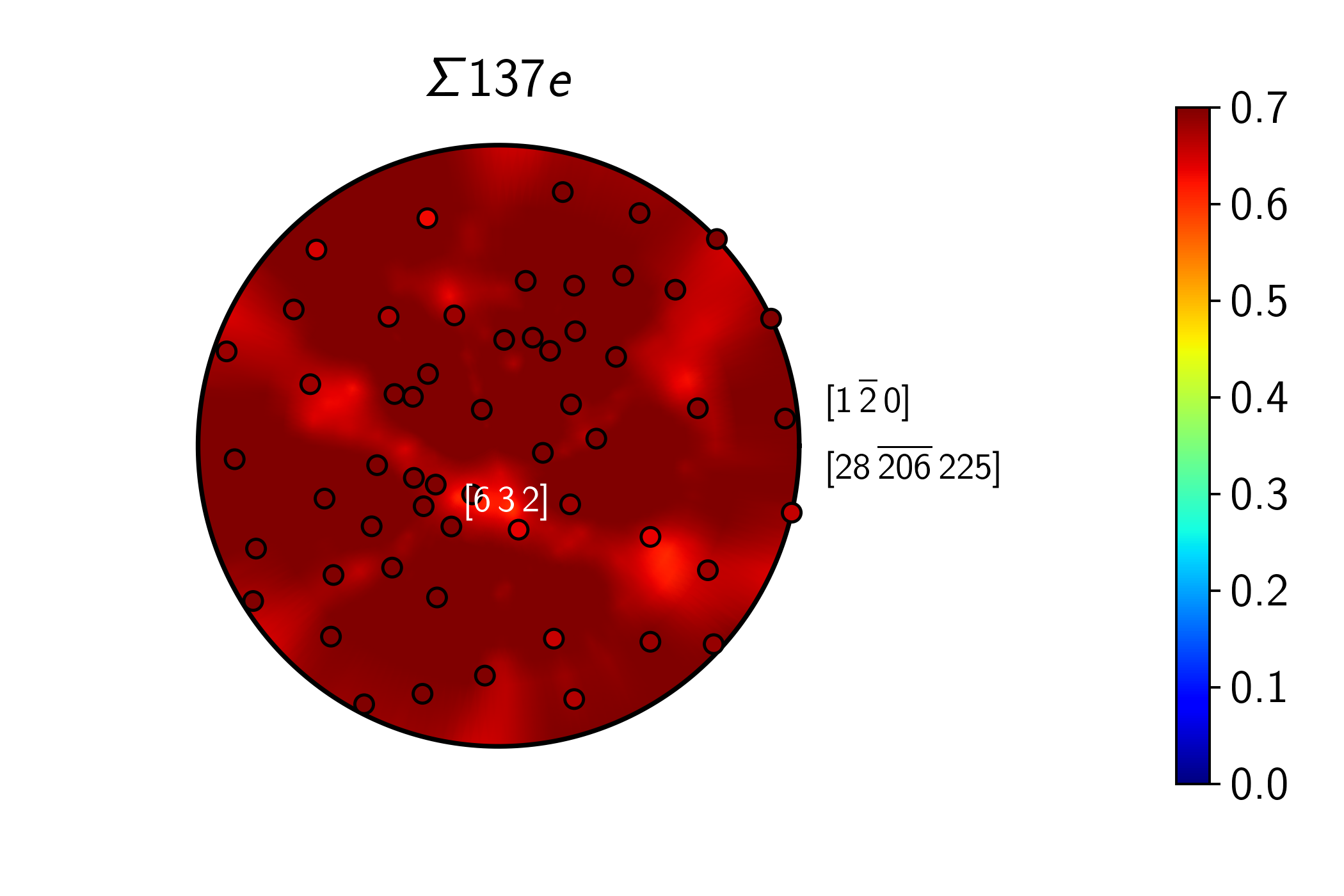} 
  \caption{$\Sigma137e - 50.0^{\circ} [6\,3\,2] - C_{i}$}\label{fig:facet_s137e}\end{subfigure}%
  \begin{subfigure}{0.24\linewidth}\includegraphics[height=\height cm,clip,trim=\tL cm \tB cm \tR cm \tT cm]{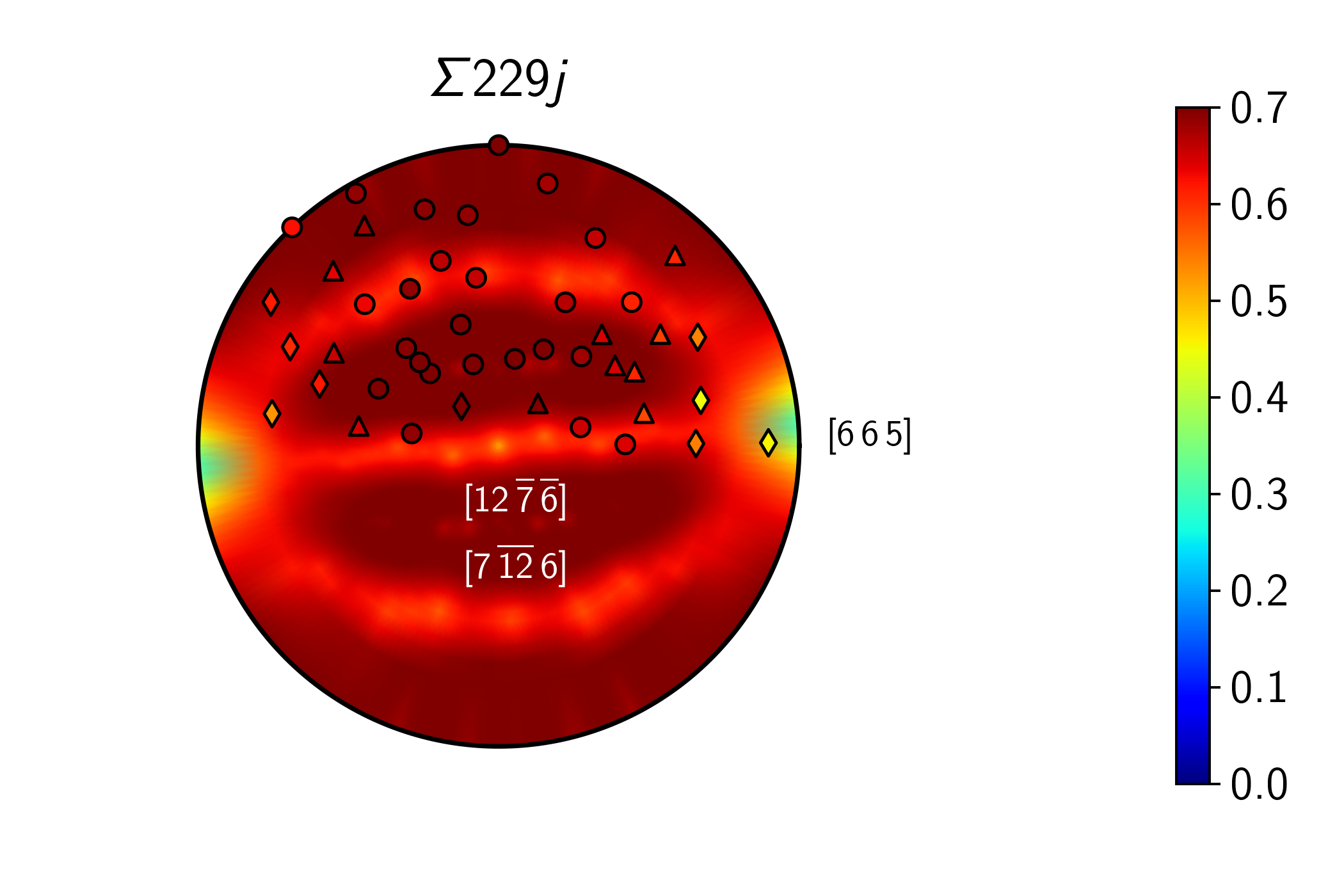} 
  \caption{$\Sigma229j - 54.8^{\circ} [6\,6\,5] - C_{2h}$}\label{fig:facet_s229j}\end{subfigure}%
  \begin{subfigure}{0.24\linewidth}\includegraphics[height=\height cm,clip,trim=\tL cm \tB cm \tR cm \tT cm]{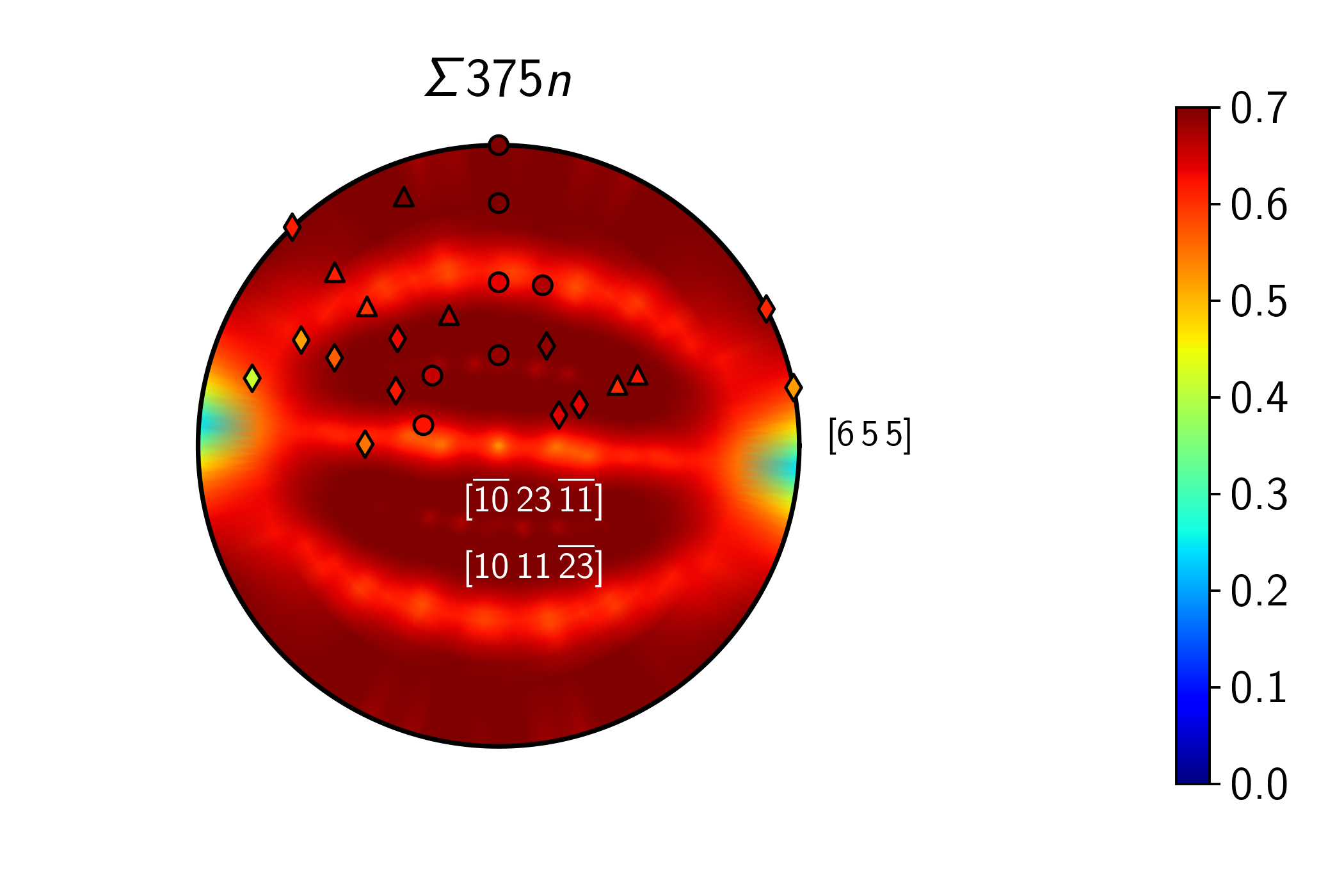} 
  \caption{$\Sigma375n - 57.2^{\circ} [6\,5\,5] - C_{2h}$}\label{fig:facet_s375n}\end{subfigure}%
  \begin{subfigure}{0.04\linewidth} \ 
  \end{subfigure}

  \caption{LM vs MD for a selection of fundamental zones. The continuous polar plot is the LM data, and the discrete points are MD. Colors correspond to the same values shown in \Cref{fig:parityplot}. Circular glyphs indicate no faceting predicted by LM, diamond glyphs indicate second order faceting predicted by LM, triangular glyphs indicate third order faceting predicted by LM.
  2-faceting and 3-faceting are considered favorable if they lower the energy by at least 10\% below the unrelaxed energy. Each of the subplots is labeled with the CSL $\Sigma$ number, the angle and axis of rotation, as well as the Sch\"onflies point group of the CSL.}
  \label{fig:polarplot}
\end{figure}

One of the key capabilities of LM is the ability to predict general trends in grain boundary space.
Grain boundary energy is characterized by the presence of deep wells (``cusps''), which are usually sharp (discontinuous derivative) at the minimum.
The presence of cusps is known to be a strong driver of boundary behavior, and it is essential that energy models for boundary energy accurately report if a certain boundary is near to a cusp.
So, while the actual prediction of the energy magnitude may be off by up to $25\%$, this is often less important than capturing the overall trend.

To explore LM predictions in general grain boundary space, we compared trends of LM and molecular dynamics in individual boundary plane fundamental zones \cite{Patala.2013.PhilosophicalMagazine,Homer:2015ie}.
The atomistic dataset considered 150 distinct fundamental zones, each with varying numbers of points depending on the symmetry of the zone.
Twelve representative zones are included here (\cref{fig:polarplot}), and the remaining 138 visualizations are included in the supplementary data\footnote{\url{fundamental_zone.tar.gz}}.
Each polar plot shows the stereographic projection of the LM energy, as defined over the unit 2-sphere, onto the unit 2-ball.

Along with the continuous LM data, discrete points from the atomistic dataset are plotted on top.
The color of each point corresponds to the value of the measured energy, scaled by the factor determined in \cref{sec:grain_boundary_energy}, so that matching colors corresponds to similar energy values.
The shape of each point corresponds to the faceting of the boundary {\it as predicted by LM}: circular points are predicted to have no facets, square points to two-faceted boundaries (when reducing energy by 10\% or more of unrelaxed) , and triangular points to three-faceted boundaries (when reducing energy by 10\% or more of unrelaxed).
A full discussion of how the LM-predicted faceting compares with that actually observed in MD will follow in the next section.

The top row of \cref{fig:polarplot} has GBs that share the $[1\,0\,0]$ disorientation axis and increase in disorientation angle from left to right. 
One can see the expected $D_{4h}$ symmetry and the general trend of increased energy with increased disorientation angle in both LM and atomistic data. 
The symmetric twist GBs about the disorientation axis (center point in the plots) consistently have the lowest energy energy in the atomistic data, but this particular location increases in energy in the LM results with increased disorientation angle. 
Of curious note in \cref{fig:facet_s265a,fig:facet_s61a} are the energy cusps along the 45 degree axes about midway between the center and edge. 
The MD data points don't show a dramatic change near the cusps but the data points are generally closer to the cusp energy than the energy predictions surrounding the cusp. 
These larger region of low energies in the MD simulations are consistent with other MD simulations of $[1\,0\,0]$ disorientation axis GBs in nickel \cite{EricksonHomer:2020hd}.

The second row of \cref{fig:polarplot} has GBs that share the $[1\,1\,1]$ disorientation axis and increase in disorientation angle from left to right. 
The LM plots all possess the expected symmetry and both LM and atomistic data feature the lowest energy at the symmetric twist GBs about the disorientation axis (center point in the plots). 
The low disorientation angle LM plot (\cref{fig:facet_s111a}) again shows cusps midway between the center and edge of the circle. These cusps are not matched by the atomistic data and the atomistic data in the surrounding points is lower than the LM prediction. On the other hand, the six-fold cusps in \cref{fig:facet_s61d,fig:facet_s3} are matched by the MD data, particularly in the case of \cref{fig:facet_s3}. 
Interestingly, these six-fold cusps are strong enough to appear in the LM predictions of CSLs somewhat close to the $\Sigma3$. 
This can be seen in \cref{fig:facet_s229j,fig:facet_s375n}, which, because of the lower $C_{2h}$ symmetry, are rotated by 90\textdegree relative to the $\Sigma3$ plot (\cref{fig:facet_s3}). 
The atomistic data in these near $\Sigma3$ GB plots appear to follow the trends but with less dramatic cusps.

The final two plots, \cref{fig:facet_s201a,fig:facet_s137e} represent a low-angle and high-angle boundary, respectively. 
The low-angle boundary in \cref{fig:facet_s201a} illustrates how the symmetric twist is not the lowest energy in  $[1\,1\,0]$ disorientation axis boundaries. 
This matches trends illustrated in \cite{Homer:2015ie}. 
Interestingly, the cusp along the horizontal axis matches the low energy of the MD simulation while the cusp along the vertical axis does not. 
The high-angle GB in \cref{fig:facet_s137e} has little indication of cusping and just generally has high energy, as indicated by both the LM and MD data.

Overall, the LM appears to give predictions that match the general trends present in the atomistic data. 
Predictions of LM cusps appear to be more strongly predicted than are evident in the atomistic data and LM also appears to miss a few cusps. 
Nonetheless LM give helpful evidence for where low-energy cusps might occur. 
The biggest differences between LM predictions and atomistic data emerge in the low-angle GBs. 
In these cases relaxed LM predictions (\cref{fig:disorientation}) predict lower, and therefore more accurate energies. 
It is known that low-angle GBs exhibit discrete dislocation arrays, after the Read-Shockley model, which is ``faceting'' after a manner and suggests why the predictions for LM get better when relaxation is allowed. 
In short, while the match is not perfect, LM provides predictions that are useful in exploring energetics in the 5D space, which can later be examined in select locations with more accurate methods.

\subsection{Lattice matching relaxation behavior} \label{sec:relaxation}

\begin{figure}
  \begin{subfigure}{0.48\linewidth}
  \includegraphics[height=5.5cm,clip,trim=0cm 0cm 0cm -1.5cm]{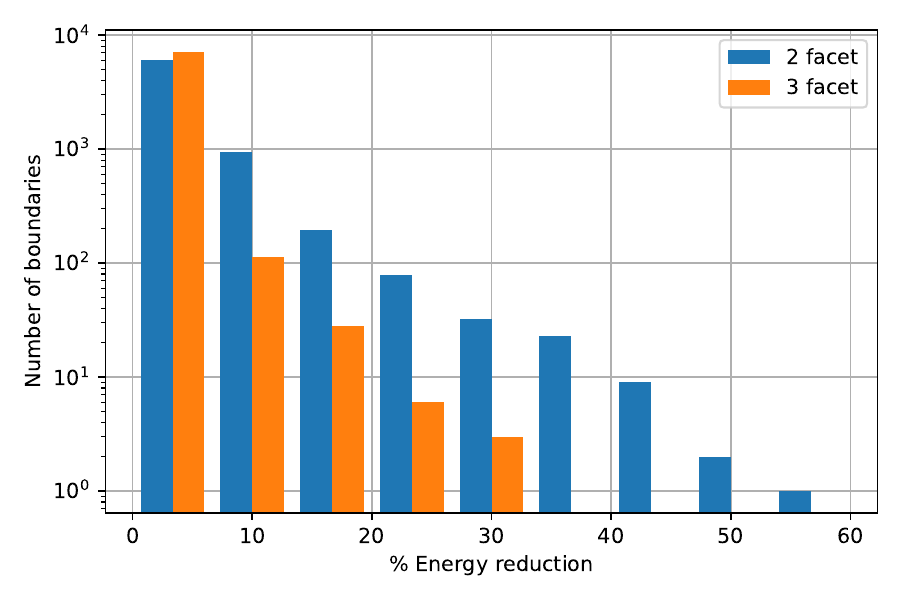}
  \caption{Facet distribution with respect to total energy reduction. Note the logarithmic scale on the ordinate.}
  \label{fig:faceting_histogram}
  \end{subfigure}\hfill
  \begin{subfigure}{0.48\linewidth}
  \includegraphics[height=5.5cm]{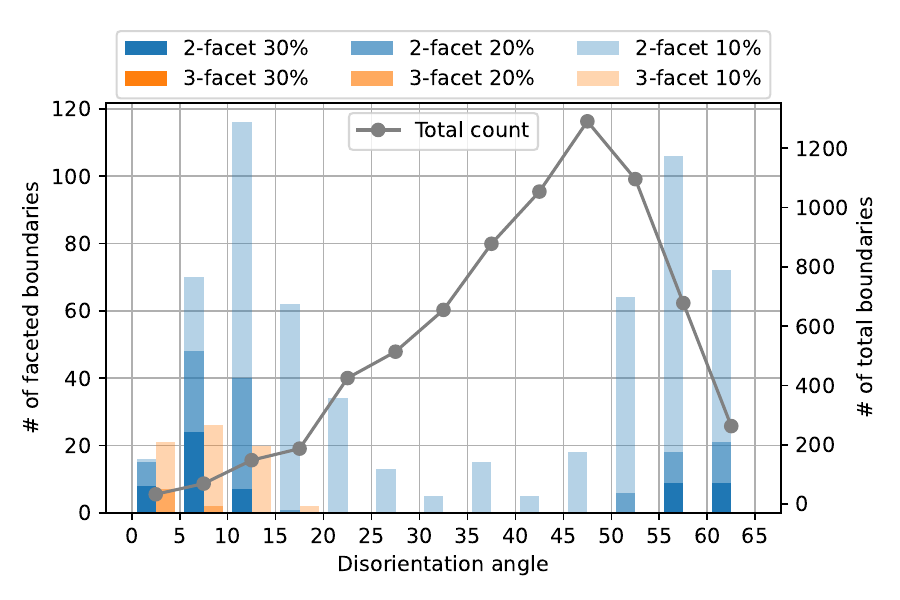}
  \caption{Facet distribution with respect to misorientation angle, and total number of boundaries (gray line)}
  \label{fig:facets_vs_disangle}
  \end{subfigure}
  \caption{Statistics of faceted boundaries as predicted by LM}
\end{figure}

\begin{figure}
  \centering
  \includegraphics[width=\linewidth]{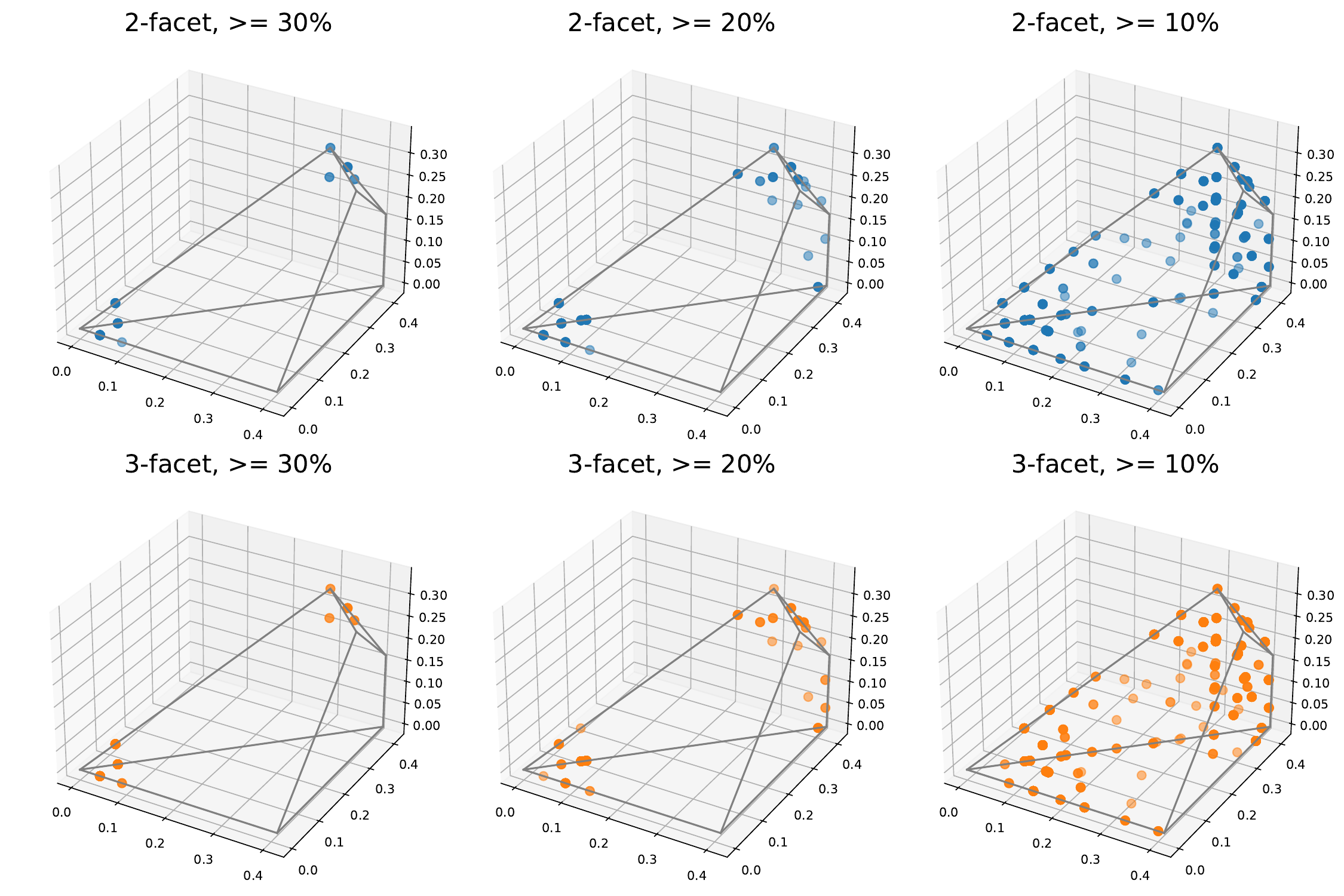}
  \caption{Plot of boundaries in the Rodrigues-Frank space with 30\% reduction (left), 20\% reduction (center), 10\% reduction (right), for second order (top, blue) and third order (bottom, gold) faceting.}
  \label{fig:facet_misorientation_space}
\end{figure}

The facet agorithm as discussed in the theory section, and as apparent from previous results, has a non-trivial effect on the accuracy of the LM model.
In this section, we quantify and explore the prevalence of LM energy reduction through faceting.
When a boundary relaxes its energy by faceting, it simultaneously lowers its energy by reorienting the boundary to more favorable orientations, and lowers its energy by creating additional boundary surface area.
In real boundaries, the junction created between facets incurs an additional energetic cost, which will likely prohibit faceting except when the energy reduction is very large.
LM does not account for this cost, and will predict faceting even if the energy reduction is marginal.
Consequently, the ``percentage of energy reduced by faceting'' is a reliable indicator for when we expect LM facet predictions to be realized in atomistic simulations.
This is especially true for the present dataset, which as mentioned earlier, was specifically designed to suppress faceting.

Of the boundaries that exhibit faceting, the vast majority accomplished an energy reduction of $30\%$ or less.
(This is reflected in \cref{fig:faceting_histogram}, where it is important to note that a logarithmic scale is used for the ordinate.)
The energy reduction measured for two-faceting is based on the original, unrelaxed energy; the energy reduction measured for three-faceting is based on the difference between the three-faceting and the two-faceting energy reductions.
Even though most faceting produced a relatively small energy reduction, there were several boundaries with an appreciable energy reduction for both two-facet and three-facet relaxation.

There is a distinct correspondence between disorientation angle and facet likelihood (\cref{fig:facets_vs_disangle}; note there is no log scale used here).
Faceting appears to occur predominantly in the 5--20$^\circ$ and 50--60$^\circ$ disorientation ranges.
Three-faceting occurs almost exclusively in low angle boundaries.
The reason for this localization is the presence of two strong cusps: the ``no boundary'' or ``near-no-boundary'' cusp for low disorientations, and the coherent $\Sigma3$ twin at high disorientations.
As stated before, small-reduction faceting does not always correspond to visble faceting in atomistic simulations, although it can be reflective of the boundary's microstructure (e.g. the Read-Shockley type structure of low-angle GBs).
This can be seen even more clearly by plotting the faceted boundaries in Rodrigues-Frank space (\cref{fig:facet_misorientation_space}), where faceted boundaries tend to localize at or near the no-boundary or the coherent twin boundary. 
In terms of quantity, the number of boundaries predicted to exhibit three-faceting at high disorientation angles is small (\cref{fig:facets_vs_disangle}).However, as we shall see in the next section, these three-facets are important for $\Sigma3$ GBs.

\subsection{Comparison of predicted and observed facet morphology}\label{sec:results_faceting}

\begin{figure}
  \centering
  \includegraphics[width=\linewidth]{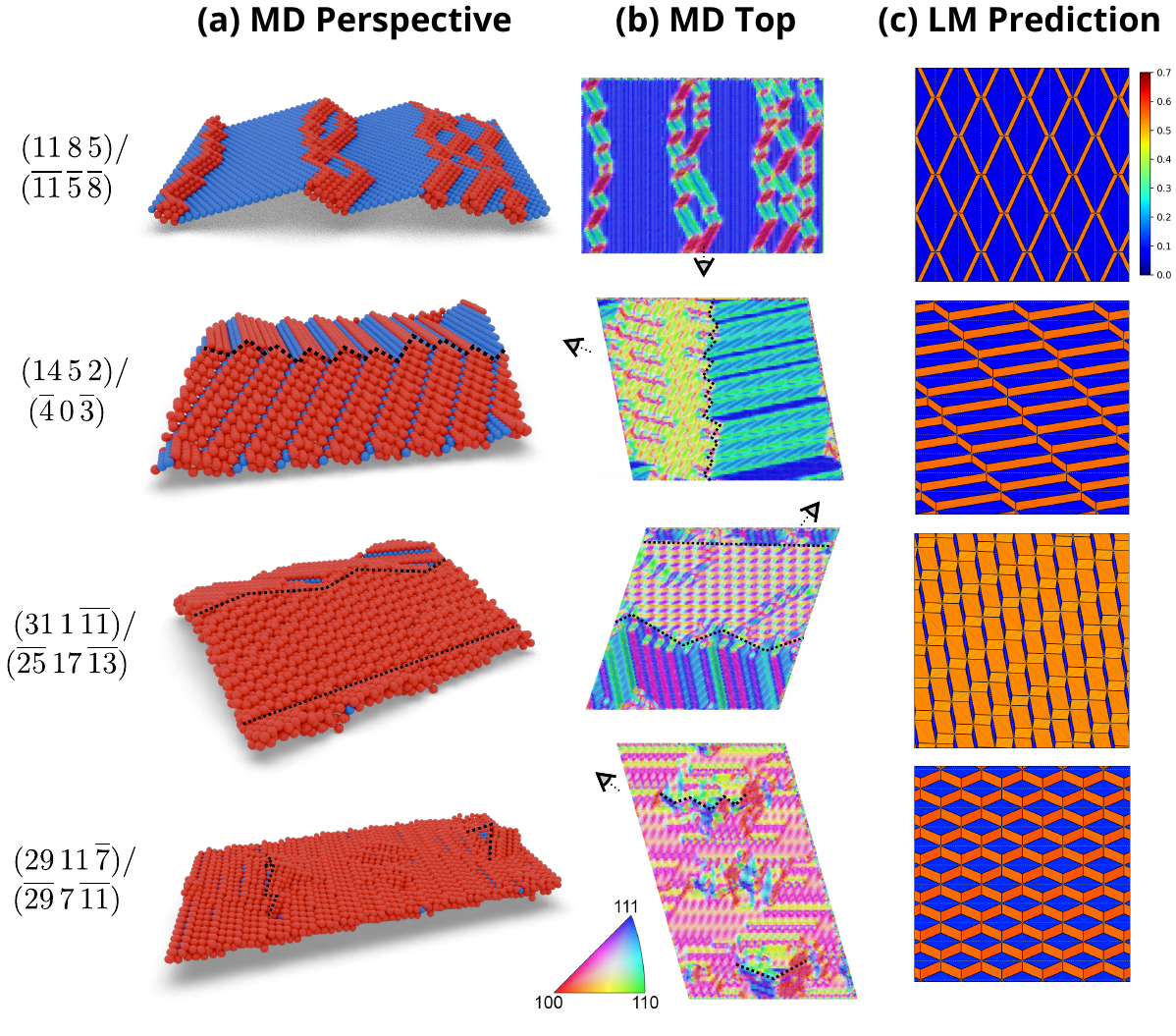}
  \caption{(a) Atomic structures of GB atoms colored by CNA (Blue = HCP atoms, Red = no crystal structure). (b) Surface mesh of GB atoms colored by the standard stereographic triangle normal to each mesh face based on the inverse pole figure coloring included. The point-of-view icon (\includegraphics[height=5pt]{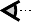}) indicates the perspective in the 3D view. (c) LM facet patterns also colored by the normal based on the inverse pole figure coloring included.}
  \label{fig:facet_comparison}
\end{figure}

To compare the facet morphology between the atomistic data and LM predictions, we apply visual and computational methods.
We focus our efforts on $\Sigma3$ GBs because of the strong drive to facet with coherent twin segments.
In the plot of energy as function of boundary plane normal (\cref{fig:polarplot}) it can be seen that the coherent twin $\hkl(111)$ and  $\hkl{112}$ normals have the lowest energy and would likely be the most ideal candidates for faceting.

In the dataset comprised of non-thermalized GBs, there are 13 $\Sigma3$ GBs, including the coherent twin.
Of these, only 1 GB exhibits considerable faceting, since GB creation process was designed not to encourage faceting.
In this work, additional annealing was carried out on these GBs to enable the GBs to form facets.
However, despite increasing the area of the GBs and annealing the structures at elevated temperatures, there was limited change to the structures in terms of the degree of faceting present.
The atomic structure of 4 $\Sigma3$ GBs is illustrated in \cref{fig:facet_comparison}a.
(Visualization and rendering of the 3D atomic structures was performed using blender \cite{blender}.)
The atoms are colored by CNA with blue atoms having HCP or twin-like local environments and red atoms having no crystal structure; all atoms with FCC structure have been removed.
It can be seen that the number of HCP or twin-like local environments decreases from top to bottom. 

To illustrate the facet normal directions, we fit the GB atoms with a mesh in OVITO \cite{Stukowski:2010ky} and color the mesh by the mesh surface normals in \cref{fig:facet_comparison}b. 
Note that these are colored by the standard stereographic triangle. 
The fraction of the GB area with $\left(1\,1\,1\right)$ normals decreases from top to bottom, but all of these structures are altered compared with the structure of the GB prior to annealing. 
\Cref{fig:facet_comparison}c plots the facet pattern predicted by LM. 
Patterns for all LM boundaries that exhibit 3-faceting are included as supplementary data.\footnote{\url{facet_pattern.tar.gz}}
The match between the facet patterns for the $\left(11\,8\,5\right)/\left(\bar{11}\,\bar{5}\,\bar{8}\right)$ GB are quite good. 
However, it can be seen that the jagged facets between the coherent twins don't quite reach $\{1\,1\,2\}$ boundary plane normals, which might conceivably have occured if the boundary were much larger or relaxed for a longer period. 
The match between the facet patterns for the $\left(14\,5\,2\right)/\left(\bar{4}\,0\,\bar{3}\right)$ GB is less clear, but in the few places where small $\left(1\,1\,1\right)$ planes can be observed, the orientation and placement of the facets appears to match. 
However, there is a large patch where the surface has numerous normals (yellow, pink, green) where the atoms don't adopt any particular facet structure. 
GBs $\left(31\,1\,\bar{11}\right)/\left(\bar{25}\,17\,\bar{13}\right)$ and $\left(29\,11\,\bar{7}\right)/\left(\bar{29}\,7\,\bar{11}\right)$ both also have a large region without any particular facet structure. 
The other regions in the GBs appear to be moving towards some sort of faceted structure, but have not yet achieved anything significant. 
For example, in the $\left(31\,1\,\bar{11}\right)/\left(\bar{25}\,17\,\bar{13}\right)$ GB in \cref{fig:facet_comparison}b an alternating pattern of $(1\,1\,1)$ and $\{1\,1\,2\}$ facets is present. 
In the $\left(29\,11\,\bar{7}\right)/\left(\bar{29}\,7\,\bar{11}\right)$ GB  several pyramid like structures are emerging. 
One of the pyramidal faces in \cref{fig:facet_comparison}a has small regions of HCP atoms, which would have a $(1\,1\,1)$ normal, but due to the small size and meshing are not colored as $(1\,1\,1)$ facets. 

The nascent facets in the bottom 3 GBs bear some resemblance to the facet patterns in \cref{fig:facet_comparison}c, but it is too early to be sure. 
In short, it appears that these system are slowly restructuring into the low energy facets, but they are limited in their ability to do so by the annealing time and the constraints of the lengthscale and the periodic boundary conditions.

\begin{figure}
  \centering
  \includegraphics[width=\linewidth]{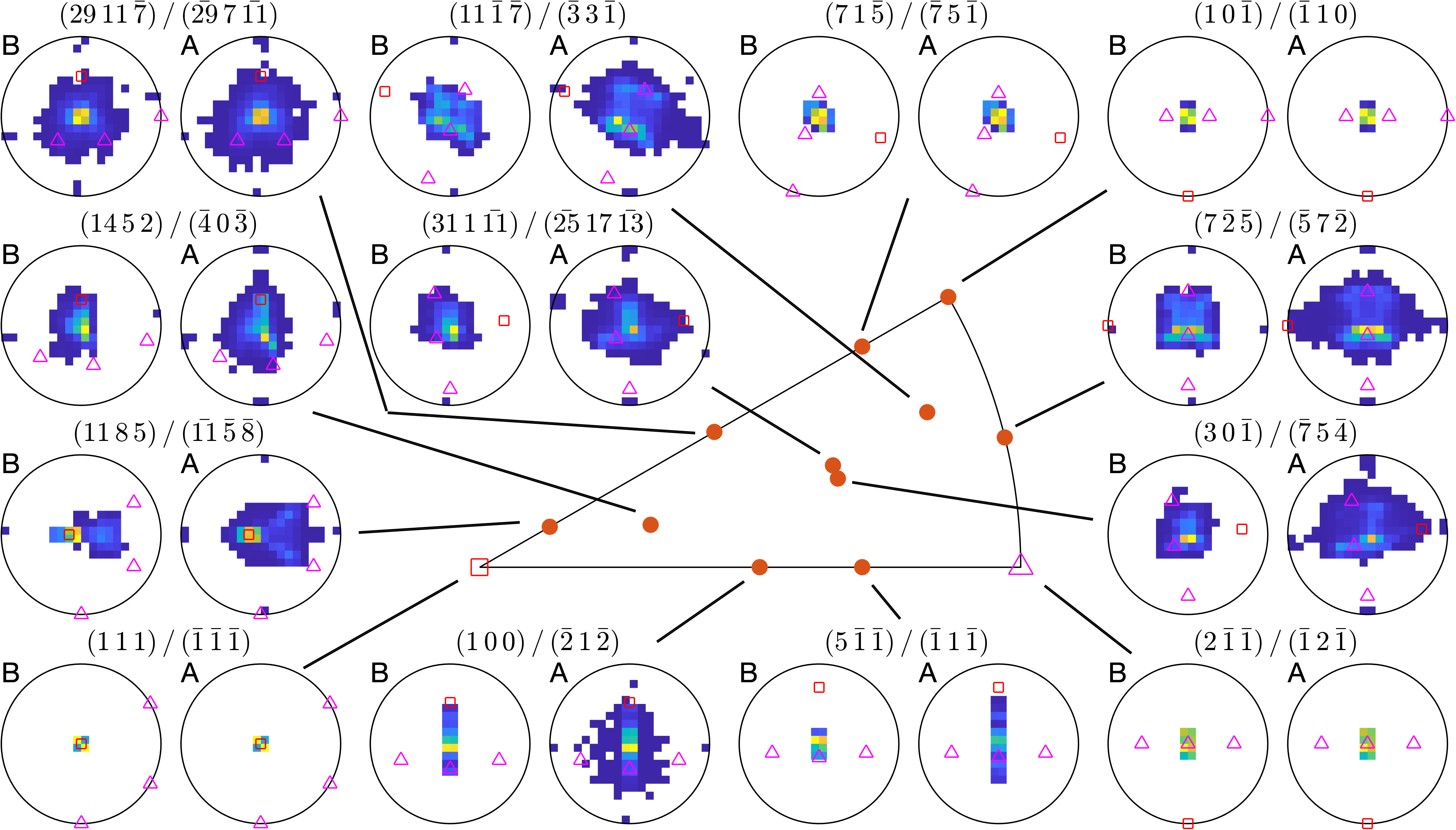}
  \caption{Plot of boundary normal distributions before (B) and after (A) annealing for the 13 $\Sigma3$ GBs. The center diagram shows the location of each GB in the $\Sigma3$ boundary plane fundamental zone. The normals for likely facets of $(1\,1\,1)$ plane and $\{1\,1\,2\}$ family of planes are denoted by a red square and magenta triangles, respectively.}
  \label{fig:facet_distributions}
\end{figure}

To better illustrate this drive towards the low energy facets, we illustrate the distribution of GB surface mesh normals for all 13 $\Sigma3$ GBs in \cref{fig:facet_distributions}. 
The distribution of normals comes from the same mesh fitting process as in \cref{fig:facet_comparison}b, except that now we plot just the distribution of normals from each GB.
\cref{fig:facet_distributions} plots the distribution of boundary plane normals before (B) and after (A) annealing.
The plots also include a red square and magenta triangles that denote the normal to the $\hkl(111)$ plane and $\hkl{112}$ family of planes, respectively. 
As noted earlier, these plane normals are the preferred facet planes for a $\Sigma3$ GB (as illustrated in  \cref{fig:polarplot}). 
Careful inspection of these normal distributions in  \cref{fig:facet_distributions} shows a tendency of the distributions to include the low energy facets, even if the system is unable to fully partition into those facets. 
It is also clear that most of the normals in the distribution have {\it not} moved towards the low energy facet positions, meaning that these GBs could perhaps lower their energy further if given sufficient time and space to facet appropriately.

Interestingly, there was negligible change in the  GB energies for all these $\Sigma3$ GBs despite the change in structures; all changes were in the range [-4,4] mJ/m$^2$. 
This is likely for the following reasons. 
First, there is a junction energy when a GB facets, which penalty affects the overall stability and lengthscale of facets that emerge \cite{Hamilton:2003:GBfaceting,Abdeljawad:2016eh}. 
Second, as noted above and as evident in \cref{fig:facet_comparison}, there are regions that do not facet in some of the GBs. 
This may be a result of the periodic boundary conditions enforced in these simulations. 
The facet geometries may not conform well to the dimensions enforced by the periodic boundaries. 
Even the facet patterns predicted by LM in \cref{fig:facet_comparison} are not setup to enforce periodic boundaries. 
Of course, periodic boundaries are a construct of the simulation methods employed and would not impact GBs in real materials, but it affects our ability to accurately simulate the nature of faceting in small simulation cells. 

In short, atomistic predictions of faceting are impacted by a number of factors, including time required to achieve lower energy structures as well as the lengthscales and boundary conditions employed. 
The lengthscales and boundary conditions not only affect the ability to achieve a pattern of facets, but also come into play by the lengthscale of the facets and frequency of facet junctions. 
Adopting facets may lower the energy but the facet junctions will raise the energy and affect the overall stability of the structure \cite{Hamilton:2003:GBfaceting,Abdeljawad:2016eh}. 
Nonetheless, in these cases LM appears to do a decent job of the likely faceting that would emerge if the GBs were given more time to achieve a faceted structure.

\section{Discussion}
\label{sec:discussion}

This work has led to a number of general insights for practitioners of molecular dynamics as well as analytic modelers of grain boundaries. 
Some of these considerations are discussed here, in the spirit of recommending best practices for future combined atomistic and theoretical efforts.

This systematic study has demonstrated that LM is useful for generating reasonably accurate GB estimations, and importantly, providing insight into the local geometry (cusps) and trends (low-angle vs high-angle GBs) given an arbirary GB character.
This makes LM a useful tool for when high-throughput GB estimates are needed, as is often the case in multiscale modeling.
On the other hand, LM users should note that LM is limited in its ability to correctly calculate the energy of high angle grain boundaries, especially as it exhibits a distinct maximum reachable energy.
It is also important to note that, while LM requires faceting relaxation in order to properly correct GB energy estimates, these facet patterns are not always reflected in the MD structures.
Therefore, care must be taken when considering LM predictions of facet morphology and comparison to MD is advisable.

From the atomistic perspective, LM can serve as a supplement that offers a quick estimate of GB energy and the kind of faceting that can occur.
Lengthscale limitations resulting from facet junction energy, and constraints of MD system size and periodicity, make it difficult to conduct a thorough search for faceted boundaries.
LM can be used to screen for facet-likely configurations, and to provide recommendations for the necessary system size and periodicity requirements in order for facets to exist.
LM can also provide information on the optimal facet configurations, so in some cases, MD can be used to individually determine facet boundary energies.

\section{Conclusions}\label{sec:conclusions}

Recent advances in experiments and computation provide unparalleled understanding of grain boundary (GB) properties. This work compares lattice matching (LM) and atomistic predictions of faceting and energy in GBs across the 5D space. In comparison to atomistic simulations, LM provides a very fast method to predict and explore GB energy and faceting with some sacrifice in accuracy. On the other hand atomistic simulations provide a detailed view of the GB energy and atomic structure, including any faceting, at the sacrifice of considerable computation time. Additionally, atomistic simulations can require considerable annealing time to form substantial facets. This work shows that LM provides reasonably accurate predictions as well as insights into the location of possible energy cusps and faceting structures.

\section*{Acknowledgements}
BD and BR acknowledge support from the National Science Foundation, grant number MOMS-2341922. CBO, GLWH, and ERH acknowledge support from the National Science Foundation, grant number DMR-1817321. The help of Aaron Gedeborg (BYU) is gratefully acknowledged.
This work used the INCLINE cluster at the University of Colorado Colorado Springs.
INCLINE is supported by the National Science Foundation, grant number OAC-2017917.

\section*{Declaration of Interests}
The authors have no conflicts to disclose.

\printbibliography

\end{document}